\documentclass[preprint,12pt]{elsarticle}
\usepackage{amssymb}
\usepackage{amsmath}
\usepackage{xurl}
\usepackage{lineno,hyperref}
\modulolinenumbers[5]

\journal{Annals of Physics}

\begin{document}

\begin{frontmatter}



\title{Free fall in KvN mechanics and Einstein's principle of equivalence}

\author{Abhijit Sen}
\address{Novosibirsk State University, Novosibirsk 630
090, Russia}
\ead{abhijit913@gmail.com}

\author{Shailesh Dhasmana}
\address{Novosibirsk State University, Novosibirsk 630
090, Russia}
\ead{dx.shailesh@gmail.com}

\author{Zurab K.~Silagadze}
\address{Budker Institute of Nuclear Physics and Novosibirsk State 
University, Novosibirsk 630 090, Russia}
\ead{Z.K.Silagadze@inp.nsk.su}

\begin{abstract}
The implementation of Einstein's principle of equivalence in  
Koopman-von Neumann (KvN) mechanics is discussed. The implementation
is very similar to the implementation of this principle in quantum mechanics.
This is not surprising, because KvN mechanics provides a Hilbert space
formulation of classical mechanics that is very similar to the quantum 
mechanical formalism.

Both in KvN mechanics and quantum mechanics, a propagator in a
homogeneous gravitational field is simply related with a free propagator.
As a result, the wave function in a homogeneous gravitational field in
a freely falling reference frame differs from the free wave function
only in phase. 

Fisher information, which quantifies our ability to estimate
mass from coordinate measurements, does not depend on the
magnitude of the homogeneous gravitational field, and this fact
constitutes the formulation of  Einstein's principle of equivalence,
which is valid both in quantum mechanics and KvN mechanics.

\end{abstract}

\end{frontmatter}


\section{Introduction}
The equivalence principle is at the heart of Einstein's general relativity. 
Usually one distinguishes between weak and strong principles of equivalence 
\cite{I1}. According to the weak principle of equivalence, the local 
gravitational acceleration is independent of the composition and structure of 
the matter being accelerated. The universality of free fall, which follows 
from the principle of weak equivalence, can be verified experimentally, it  
has played an important role historically and still remains the subject of 
intense theoretical and experimental research \cite{I2,I3,I4}.

Strong principle of equivalence asserts much more \cite{I2,I4}: 
\begin{enumerate}
\item The weak  principle of equivalence is valid for self-gravitating bodies 
as well as for test bodies.
\item The outcome of any local test experiment is independent of the velocity 
and of the orientation of the freely falling laboratory in which the 
experiment is performed.
\item The outcome of any local test experiment is independent of where and 
when it is performed.
\end{enumerate}
In contrast to the weak principle of equivalence, the strong principle of 
equivalence is somewhat ambiguous like the correspondence principle between 
classical and quantum mechanics \cite{I5}. For example, the second facet of
the strong principle of equivalence, the local Lorentz invariance, is 
heuristically useful for generalizing physical laws tentatively known in the 
absence of gravity to the case when gravity is present. However, when the 
second or higher derivatives occur in the physical law, an ambiguity arises
since the corresponding covariant derivatives do not commute. This  ambiguity 
is analogous to the operator ordering ambiguity in the quantum-classical
correspondence \cite{I5}. 

Another possible source of confusion in the formulation of the strong  
principle of equivalence is the term ``local''. From a practical point of 
view, this, of course, means that the laboratory should be small enough and 
the duration of the experiment short enough to neglect the effects of 
spacetime curvature. However, whether one can neglect the effects of the 
spacetime curvature depends also on the sensitivity of the laboratory 
apparatus.

It is not surprising, therefore, that several non-equivalent formulations
of the principle of equivalence appeared in the literature in order to make 
the principle more accurate \cite{I6}. Moreover, there is no universal 
agreement on the meaning and implementation of the equivalence principle 
within a relativistic framework (see references cited in \cite{I6A}). Einstein 
himself always insisted on the fundamental importance of this principle to 
the general theory of relativity. This insistence created a puzzle for 
philosophers and historians of science \cite{I7}, because to maintain 
Einstein's view that an accelerated laboratory in the absence of gravity is 
strictly equivalent to a laboratory at rest in a gravitational field, and thus 
acceleration, like velocity, is a relative quantity, one should assume 
sufficiently small (in fact, infinitesimal) spacetime regions and limit 
attention to sufficiently elementary phenomena, because otherwise, due to the 
presence of curvature, one can detect the difference between the two 
situations \cite{I6}.
 
It is expected that the weak principle of equivalence will be violated in 
quantum mechanics if it is understood as the independence of the dynamics of 
a test body on its mass in an external gravitational field. A simple argument 
showing that not all consequences of the equivalence principle in the 
classical domain are valid in the quantum domain is as follows 
\cite{I8A,I8}.

Imagine a bound quantum system of neutron and electron in which the 
gravitational interaction is responsible for the bound state formation. One 
can easily see after some calculations that energy eigenvalues depend on 
the electron mass $m$ (more precisely, are proportional to $m^3$). Therefore,
the frequency of transition from one energy eigenstate to another will also 
depend on $m$. These transition  frequencies, in principle, can be 
experimentally observed, measured and used to find the mass of the electron. 
In contrast, in classical theory, the dynamics of a satellite orbiting 
a massive body does not depend on the mass of the satellite and, therefore,
cannot be used to determine its value.
   
The crux of the problem, when we try to extend the weak principle of 
equivalence in quantum domain, lies in the fact that, in contrast to the
Newton equation, when considering  the behavior of a quantum particle in 
an external gravitational field, the mass of the particle does not drop from 
the corresponding Schr\"{o}dinger equation.

However, the same thing happens with a free quantum particle too and this 
circumstance offers a way to understand and generalize Einstein's principle of 
equivalence in the quantum domain \cite{I9}. Einstein's principle of 
equivalence is the assertion that a state of rest in a homogeneous 
gravitational field is physically indistinguishable from a state of uniform 
acceleration in a gravity-free space. And in this form, the principle is valid 
in the quantum domain too, since it can be shown that the solution of the 
Schr\"{o}dinger equation for a particle in a homogeneous gravitational field 
$\vec{g}$, as viewed from the reference frame undergoing a constant 
acceleration $\vec{a}$, coincides to the free particle wave function up to 
a phase, if  $\vec{a}=\vec{g}$ \cite{I9,I10}. That is, in quantum mechanics, 
like classical general relativity or Newtonian gravity, all effects of gravity 
are cancelled in a freely falling reference frame.

It was shown by Koopman \cite{I11} and von Neumann \cite{I12} long ago that
classical mechanics can be formulated in the language of Hilbert spaces, in 
the way which resembles operator formulation of quantum mechanics. The purpose 
of the present article is to investigate free fall in the framework of 
Koopman-von Neuman  (KvN) mechanics and show that Einstein's principle of 
equivalence is implemented in KvN mechanics in much the same way as in ordinary
quantum mechanics.
 
We tried to make the paper as self-contained as possible, providing 
the necessary pedagogical details where, in our opinion, they were needed.
As a result, the article contains a significant amount of review material
which is organized into several appendices.

\section{Einstein's principle of equivalence and quantum mechanics}
The  Schr\"{o}dinger equation in the inertial frame $S$ (where the 
gravitational potential is present) is given by
\begin{equation}
i\hbar \frac {\partial \psi \left( x,t\right) }{\partial t}=-\frac {\hbar ^{2}}
{2m}\frac {\partial ^{2}\psi \left( x,t\right) }{\partial x^{2}}+mgx\,
\psi (x,t)=\left (\frac{{\hat p}^2}{2m}+mg\hat x \right )\psi (x,t),
\label{eq27}
\end{equation}
where $\hat x$ and $\hat p$ operators satisfy the canonical commutation 
relation $[\hat x,\,\hat p]=i\hbar$. The formal solution of this 
Schr\"{o}dinger equation has the form
\begin{equation}
\psi (x,t)=e^{-\frac{i}{\hbar}\left (\frac{{\hat p}^2}{2m}+mg\hat x \right )t}\psi (x,0),
\label{eq28}
\end{equation}
where $\psi (x,0)$ is the wave function of the initial state. The true 
solution of (\ref{eq27}) then can be found by disentangling noncommutative 
operators in this formal solution (such operator  methods were pioneered by 
Richard Feynman \cite{29,30}).

To disentangle the exponential operator in (\ref{eq28}), one can use the 
``left oriented'' version of the generalized Baker-Campbell-Hausdorff 
formula \cite{31} 
\begin{equation}
e^{\lambda (\hat A+\hat B)}=e^{\frac{\lambda^3}{3!}\left (2[\hat B,\,[\hat A,\,\hat B
]]+[\hat A,\,[\hat A,\,\hat B]]\right )}\,e^{\frac{\lambda^2}{2!}[\hat A,\,\hat B]}\,
e^{\lambda \hat B}\,e^{\lambda \hat A}.
\label{eq29}
\end{equation}
This second order formula is valid if
$$[\hat A,\,[\hat A,\,[\hat A,\,\hat B]]]=
[\hat B,\,[\hat B,\,[\hat A,\,\hat B]]]=[\hat A,\,[\hat B,\,[\hat A,\,\hat B
]]]=[\hat B,\,[\hat A,\,[\hat A,\,\hat B]]]=0.$$
In particular, if we take $\lambda =-\frac{i}{\hbar}t$, $\hat A=
\frac{{\hat p}^2}{2m}$, $\hat B=mg\hat x$, then
\begin{equation}
[\hat A,\,\hat B]=-ig\hbar\hat p,\;\;[\hat A,\,[\hat A,\,\hat B]]=0,\;\;
[\hat B,\,[\hat A,\,\hat B]]=mg^2\hbar^2,
\label{eq30}
\end{equation}
and applying (\ref{eq29}) to (\ref{eq28}), we obtain
\begin{equation}
\psi (x,t)=e^{\frac{i}{\hbar}\frac{mg^2t^3}{3}}\,e^{\frac{i}{\hbar}\frac{gt^2}{2}\,
\hat p}e^{-\frac{i}{\hbar}\,mg\hat x t}e^{-\frac{i}{\hbar}\,\frac{{\hat p}^2}
{2m}\,t}\,\psi (x,0).
\label{eq31}
\end{equation}
It is clear that $e^{-\frac{i}{\hbar}\,\frac{{\hat p}^2}{2m}\,t}\,\psi (x,0)=\phi(x,t)$
gives  the wave function of a free particle (that is, $\phi(x,t)$ is a solution
of the Schr\"{o}dinger equation with zero potential). On the other hand, 
$e^{\frac{i}{\hbar}a\hat p}$ is a spatial translation operator and  for  any 
arbitrary well behaved function $f(x)$, we have $ e^{\frac{i}{\hbar}a\hat p}f(x)=
f(x+a)$. Therefore
$$e^{\frac{i}{\hbar}\frac{gt^2}{2}\,\hat p}e^{-\frac{i}{\hbar}\,mg\hat x t}\,\phi(x,t)=
e^{-\frac{i}{\hbar}\,mg\left(x+\frac{gt^2}{2}\right) t}\,\phi\left (x+\frac{gt^2}{2},
t\right ),$$
and (\ref{eq31}) takes the form
\begin{equation}
\psi \left( x,t\right) = e^{-\frac {imgt}{\hbar }\left( x+
\frac {gt^{2}}{6}\right)}\,\phi \left( x+\frac {gt^{2}}{2},t\right).
\label{eq32} 
\end{equation}
Expressing $x$ in terms of $x^\prime$ through (\ref{eq19}), we see that the phase
in the above exponential prefactor exactly equals, as expected, to 
$\Lambda/\hbar$, with $\Lambda$ specified in (\ref{eq20}).
 
To better understand the physical meaning of (\ref{eq32}), let's make the 
change of variables (\ref{eq19}), which, with a passive interpretation,
corresponds to  a transition to a freely falling reference frame, in 
the Schr\"{o}dinger equation (\ref{eq27}). Then the following picture emerges 
\cite{I10,32,33}.

It follows from (\ref{eq19}) that
$$\frac{\partial x^\prime}{\partial x}=1,\;\;\frac{\partial x^\prime}{\partial t}
=gt=gt^\prime,\;\;\frac{\partial t^\prime}{\partial x}=0,\;\;
\frac{\partial t^\prime}{\partial t}=1.$$
Therefore,
\begin{equation}
\frac{\partial }{\partial t}=\frac{\partial t^\prime}{\partial t}\,
\frac{\partial }{\partial t^\prime}+\frac{\partial x^\prime}{\partial t}\,
\frac{\partial }{\partial x^\prime}=\frac{\partial }{\partial t^\prime}+
gt^\prime \frac{\partial }{\partial x^\prime},\;\;
\frac{\partial }{\partial x}=\frac{\partial t^\prime}{\partial x}\,
\frac{\partial }{\partial t^\prime}+\frac{\partial x^\prime}{\partial x}\,
\frac{\partial }{\partial x^\prime}=\frac{\partial }{\partial x^\prime},
\label{eq33} 
\end{equation}
and the Schr\"{o}dinger equation (\ref{eq27}) in new coordinates takes the form
\begin{equation}
i\hbar\left (\frac{\partial }{\partial t^\prime}+gt^\prime \frac{\partial }
{\partial x^\prime}\right )\psi(x^\prime,t^\prime)=-\frac{\hbar^2}{2m}\,
\frac{\partial^2 \psi(x^\prime,t^\prime)}{\partial x^{\prime\, 2}}+mg\left (x^\prime-
\frac{1}{2}gt^{\prime\,2}\right )\psi(x^\prime,t^\prime).
\label{eq34} 
\end{equation}
Let's try a solution in the form
\begin{equation}
\psi(x^\prime,t^\prime)=e^{\frac{i}{\hbar}\Lambda(x^\prime,t^\prime)}\phi(x^\prime,t^\prime),
\;\;\Lambda(x^\prime,t^\prime)=\frac{1}{3}mg^2t^{\prime\,3}-mgx^\prime t^\prime.
\label{eq35} 
\end{equation}
Note that $\Lambda(x^\prime,t^\prime)$ is precisely the function specified in 
(\ref{eq20}). Substituting (\ref{eq35}) into  (\ref{eq34}), we get after some 
straightforward calculation
\begin{equation}
i\hbar\frac{\partial \phi(x^\prime,t^\prime)}{\partial t^\prime}=
-\frac{\hbar^2}{2m}\,\frac{\partial^2 \phi(x^\prime,t^\prime)}
{\partial x^{\prime\, 2}}.
\label{eq36} 
\end{equation} 
Therefore $\phi(x^\prime,t^\prime)$ satisfies the Schr\"{o}dinger equation for 
a free particle.

Now the physical meaning of (\ref{eq32}) is clear: in a freely falling 
reference frame the wave function of a particle, not subject to 
non-gravitational forces, differs from the wave function of a free particle 
only by a phase factor. This circumstance can be considered as a 
quantum-mechanical embodiment of  Einstein's principle of equivalence.

However, there is a subtlety. Although the overall phase of the wave function
is unobservable, the relative phase between its two components does have 
observable consequences, and the phase in (\ref{eq32}) depends on the position 
of the particle. Due to this position dependence, if a beam of quantum 
particles is coherently split into two parts in a gravitational field and then 
recombined, the interference effects may appear that will be mass dependent
\cite{I8A}.

A beautiful experiment of this type was performed by Colella, Overhauser and 
Werner \cite{34} in which the quantum-mechanical phase shift of neutrons 
induced by Earth's gravitational field was observed. 

In this experiment, an almost monoenergetic horizontal beam of thermal neutrons 
was coherently  divided into two secondary beams using Bragg reflection in the 
first silicon crystalline wafer. The secondary beams move at different heights 
in the Earth's gravitational field, and then each again coherently splits in 
the second silicon slab. Two of the resulting beams are directed to the same 
point in the third silicon plate, where they overlap and interfere with each
other. Finally, the outgoing interfering beams are detected by neutron 
counters.

Assuming for simplicity the constant height difference $H$ between the 
interfering beams on the horizontal sections of their trajectories, the 
equation (\ref{eq32}) tells us that, when the beams recombine, the accumulated 
phase difference will be $\Delta\Phi=mgtH/\hbar$. But $t=L/v$, where $L$ is
the length of the parallelogram along which the beams travel, and $v=p/m=h/m
\lambda$ is the velocity of the neutron with the de de Broglie wavelength 
$\lambda$. Therefore, the expected phase difference is \cite{32}
\begin{equation}
\Delta \Phi=\frac{m^2g\lambda}{2\pi\hbar^2}\,A,\;\;A=HL.
\label{eq37} 
\end{equation} 
Of course, this simplified treatment, although it correctly conveys the main 
idea of the experiment, says nothing about the real difficulties that 
accompanied the implementation of the COW experiment, which required incredible
experimental and engineering skills in the field of neutron scattering
\cite{35}. 

The fact that the phase shift (\ref{eq37}) depends explicitly on the 
neutron mass triggered a heated discussion with  no consensus up to now on 
whether the principle of equivalence applies to quantum systems, and what form
it should take in quantum domain \cite{36}. For example, it was stated in the 
well-known textbook on quantum mechanics \cite{37} that the COW experiment 
shows that gravity is not purely geometric at the quantum level, since the 
effect depends on $m/\hbar$.

In reality, however, the COW experiment simply shows that in non-relativistic
quantum mechanics gravity appears in the Schr\"{o}dinger equation as any other
force would. And this circumstance does not compromise the geometrical 
description of gravity. In fact, it is possible to obtain the phase shift
(\ref{eq37}) by considering the non-relativistic limit of the Dirac equation in
the general relativistic Schwarzschild metric \cite{38}. 

In classical mechanics, the weak equivalence principle has three complementary 
forms \cite{39}: 1) free-fall trajectories do not depend on the masses of the 
test particles; 2) inertial and (passive) gravitational masses are equal;
3) it is always possible to locally eliminate the gravitational field by
going into a free falling reference frame.

It is clear that the first form is problematic in quantum mechanics, since
a quantum particle, in general, does not have a well-defined trajectory.
A possible quantum generalization of this form of equivalence principle, namely
that the wave function of a freely falling particle does not depend on its 
mass, is also not feasible, since the wave function does depend on the mass of 
the particle. 

However, this impossibility to give a quantum version of the first form of the 
equivalence principle does not mean, as it is sometimes claimed, that quantum 
mechanics violates the equivalence principle. The mass dependence of the 
quantum-mechanical wave function is completely natural, as the following 
reasoning shows \cite{40}.

Geodesic equation (\ref{eq14}) (its relativistic version) follows from the
variational principle $\delta S=0$, where $S=-mc\int d\tau$ is the covariant
classical action. The resulting classical trajectory is independent of the 
mass $m$, which enters in the action just as a kinematic overall 
multiplying factor. However the actual value of $S$ along classical trajectory
depends on $m$ (is proportional to it). This value of $S$ is classically 
irrelevant, since it is unobservable.  Nonetheless, it is observable in
quantum mechanics as the phase $S/\hbar$ of the wave function (in the 
semiclassical approximation). Although the values of the classical action
on different neutron paths in the COW experiment differ by a very small
amount $mgtH=mgHL/v$, this difference is not small compared to the Planck's 
constant and leads to the observed interference pattern \cite{40}.

The second and third versions of the equivalence principle retain their 
operational meaning in the quantum domain too. In fact, COW type experiments
are quantum tests of the equality of inertial and (passive) gravitational 
masses, albeit with a modest accuracy of about 1\%. Despite many improvements
to the original COW experiment, a long-standing small discrepancy of about 
1\% between theory and experiment remained. This discrepancy probably is related
to the slight intrinsic misalignments between diffracting crystals \cite{41}.

To verify that the same phase shift as in the gravitational field $\vec{g}$
is induced without gravity, but instead having the acceleration $-\vec{g}$,
Bonse and Wroblewski conducted an experiment with a horizontally oscillating
interferometer \cite{42} and measured the phase shift as a function of the 
maximum acceleration of the apparatus. The results confirmed Einstein's 
principle of equivalence in the quantum domain with an accuracy of about 4\%.

\section{Einstein's principle of equivalence and KvN  mechanics}
According to Sudarshan, any KvN system can be considered as a subsystem of a 
genuine quantum system with doubled degrees of freedom and with a certain
quantum dynamics. Therefore, it  is expected that the implementation of the 
Einstein's principle of equivalence in KvN  mechanics will be very similar 
to how it is done in quantum mechanics. This is  indeed true as shown below.

The Schr\"{o}dinger equation (\ref{eq66}) for the potential $V(x)=mgx$ 
(a classical particle of mass $m$ in a uniform  gravitational field $g$
in a negative $x$-direction) has the form
\begin{equation}
i\hbar\frac{\partial \psi(x,X,t)}{\partial t}=\left (\frac{\hat p\hat P}{m}+
mg\hat X\right )\psi(x,X,t).
\label{eq67}
\end{equation}
Its formal solution is
\begin{equation}
\psi (x,X,t)=e^{-\frac{i}{\hbar}\left (\frac{\hat p\hat P}{m}+mg\hat X \right )t}
\psi (x,X,0),
\label{eq68}
\end{equation}
where $\psi (x,X,0)$ is the wave function of the initial state. Again we
should disentangle noncommutative operators in this formal solution. For this
purpose, we use the generalized Baker-Campbell-Hausdorff formula (\ref{eq29}) 
with 
$\lambda =-\frac{i}{\hbar}t$, $\hat A=\frac{\hat p\hat P}{m}$, $\hat B=mg
\hat X$. Then
\begin{equation}
[\hat A,\,\hat B]=-ig\hbar\hat P,\;\;[\hat A,\,[\hat A,\,\hat B]]=0,\;\;
[\hat B,\,[\hat A,\,\hat B]]=0,
\label{eq69}
\end{equation}
and applying (\ref{eq29}) to (\ref{eq68}), we obtain
\begin{equation}
\psi (x,X,t)=e^{\frac{i}{\hbar}\frac{gt^2}{2}\,
\hat P}e^{-\frac{i}{\hbar}\,mg\hat X t}e^{-\frac{i}{\hbar}\,\frac{\hat p\hat P}
{m}\,t}\,\psi (x,X,0).
\label{eq70}
\end{equation}
But $e^{-\frac{i}{\hbar}\,\frac{\hat p\hat P}{m}\,t}\,\psi (x,X,0)=\phi(x,X,t)$
gives the wave function of a free particle ($\phi(x,X,t)$ is a solution
of the Schr\"{o}dinger equation  (\ref{eq67}) with $g=0$). On the other hand, 
$e^{\frac{i}{\hbar}a\hat P}$ is a spatial translation operator in the $x$-direction,
as the commutation relations (\ref{eq59}) do indicate. Therefore
$$e^{\frac{i}{\hbar}\frac{gt^2}{2}\,\hat P}e^{-\frac{i}{\hbar}\,mg\hat X t}\,\phi(x,X,t)=
e^{-\frac{i}{\hbar}\,mgX t}\,\phi\left (x+\frac{gt^2}{2},X,t\right ),$$
and (\ref{eq70}) takes the form
\begin{equation}
\psi \left( x,X,t\right) = e^{-\frac{i}{\hbar}mgXt}\,\phi \left( x+\frac {gt^2}
{2},
X,t\right).
\label{eq71} 
\end{equation}

Now we obtain this result in a different way that also parallels the 
quan\-tum-mechanical case, as considered in \cite{32}. The following passive 
transformation
\begin{equation}
t^\prime=t \;\;\; x^\prime=x+\frac{gt^2}{2} \;\;\; X^\prime=X,
\label{eq72}
\end{equation}
describes the transition to a freely falling reference frame, the origin 
of which moves in the negative $x$-direction with acceleration $g$. In the new
coordinates, the Schr\"{o}dinger equation  (\ref{eq67}) takes the form
\begin{equation}
i\hbar \left( \frac {\partial \psi }{\partial t^\prime}+gt^\prime\frac {\partial 
\psi}{\partial x^\prime}\right) =-\frac {\hbar ^{2}}{m}\frac {\partial^2\psi }
{\partial x^\prime\partial X^\prime}+mgX^\prime\psi.
\label{eq73}
\end{equation}
Let's try the solution in the form
\begin{equation}
\psi \left( x^\prime,X^\prime,t^\prime\right) =\Phi \left( x^\prime,X^\prime,t^\prime
\right) \exp{\left[ i\eta ( x^\prime,X^\prime,t^\prime\right]},
\label{eq74}
\end{equation}
where $\Phi \left(x^\prime,X^\prime,t^\prime\right)$ satisfies the free 
Schr\"{o}dinger equation 
\begin{equation}
i\hbar \frac {\partial \Phi }{\partial t^\prime}=-\frac {\hbar^2}{m} 
\frac {\partial^2\Phi}{\partial x^\prime\partial X^\prime}.
\label{eq75}
\end{equation}
Substituting (\ref{eq74}) into (\ref{eq67}) and using (\ref{eq75}), we get
\begin{eqnarray} &&
i\hbar \left[ i\frac {\partial \eta}{\partial t^\prime}\Phi +gt^\prime\left( 
\frac {\partial \Phi }{\partial x^\prime}+i\frac {\partial \eta}{\partial 
x^\prime}\Phi \right) \right]  = \\ &&
-\frac {\hbar^2}{m}\left[ i\frac {\partial \Phi }{\partial x^\prime}
\frac {\partial \eta}{\partial X^\prime}+i\frac {\partial^2 \eta}{\partial 
x^\prime\partial X^\prime} \Phi+i\frac {\partial \eta}{\partial x^\prime}
\frac {\partial \Phi }{\partial X^\prime}-\frac {\partial  \eta}{\partial 
x^\prime}\frac {\partial \eta}{\partial X^\prime}\Phi \right] +mgX^\prime\Phi. 
\nonumber
\label{eq76}
\end{eqnarray}
The coefficients of $\Phi$, $\frac{\partial \Phi}{\partial x^\prime}$ and 
$\frac{\partial \Phi}{\partial X^\prime}$ must match separately. Since we have 
$\frac{\partial \Phi}{\partial X^\prime}$ only in one place, hence 
$\frac{\partial\eta}{\partial x^\prime}=0$. Therefore, unlike the quantum 
case, the phase does not depend on the configuration space coordinate 
$x^\prime$, but may depend on the hidden coordinate $X^\prime$. Then, matching 
coefficients of $\Phi$ and $\frac{\partial\Phi}{\partial x^\prime}$, we find
\begin{equation}
-\hbar\,\frac {\partial \eta }{\partial t^\prime}=mgX^\prime,
\label{eq77}
\end{equation}
and
\begin{equation}
i\hbar gt^\prime=-i\,\frac {\hbar^2}{m}\,\frac {\partial \eta }
{\partial X^\prime}.
\label{eq78}
\end{equation}
Eq.(\ref{eq77}) can be integrated with the result
$$\eta =-\frac {mgt^\prime}{\hbar }X^\prime+h\left( X^\prime\right),$$
where $h$ is an arbitrary function which must be a constant in order 
(\ref{eq78}) to be valid. Therefore, up to this irrelevant constant in phase, 
\begin{equation}
\psi \left( x^\prime,X^\prime,t^\prime\right) =\Phi \left( x^\prime,X^\prime,t^\prime
\right) \exp \left[ -\frac {i}{\hbar }mgt^\prime X^\prime\right],
\label{eq79}
\end{equation}
which is equivalent to (\ref{eq71}). Therefore, as in the quantum case, in a 
freely falling reference frame the wave function of a particle not subject to 
non-gravitational forces differs from the wave function of a free particle 
only by a phase factor. However, in contrast to quantum mechanics, this phase
depends only on a hidden coordinate, and not on the observable position of the 
particle, and thus cannot lead to any observable interference effects.

Still another parallel with quantum-mechanical considerations is provided 
by the following observation. Let us make the following canonical 
transformation:
\begin{eqnarray} &&
\hat x=\frac {1}{\sqrt{2}}\left(\hat x_1-\hat x_2\right),\;\; \hat X=\frac{1}
{\sqrt {2}}\left(\hat x_1+\hat x_2\right), \nonumber \\ &&
\hat p=\frac {1}{\sqrt {2}}\left( \hat p_1+\hat p_2\right), \;\;\;  
\hat P=\frac {1}{\sqrt{2}}\left( \hat p_1-\hat p_2\right).
\label{eq80}
\end{eqnarray}
In new coordinates the Hamiltonian $\hat H=mg\hat X+\frac{\hat p \hat P}{m}$ 
takes the form
\begin{equation}
\hat H=\left(\frac {{\hat p}^2_1}{2m}+m\dfrac {g}{\sqrt {2}}\hat x_{1}\right) 
-\left(\frac {{\hat p}^2_2}{2m}+m\frac {\left( -g\right) }{\sqrt {2}}\hat x_2
\right)=\hat H_1-\hat H_2.
\label{eq81}
 \end{equation}
Then we have
\begin{equation}
\psi \left( x_1,x_2,t\right) =\psi_1\left(x_1,t\right) \psi_2\left(x_2,-t
\right),
\label{eq82}
\end{equation}
where $\psi_1\left(x_1,t\right)$ and $\psi_2\left(x_2,t\right)$ separately 
satisfy the Schr\"{o}dinger equation with Hamiltonians $\hat H_1$ and
$\hat H_2$ respectively. According to (\ref{eq32}) (one should be careful with 
the gravitational acceleration term which is $\frac{g}{\sqrt {2}}$ in 
$\hat H_1$, but $-\frac{g}{\sqrt {2}}$ in $\hat H_2$),
\begin{eqnarray} &&
\psi_1\left(x_1,t\right)  = \Phi_1\left(x_1+\frac{gt^2}{2\sqrt{2}}\,,t\right) 
\exp{\left[ \frac {-imgt}{\sqrt {2}\hbar }\left( x_1+\frac {gt^2}
{6\sqrt {2}}\right) \right]}, \nonumber \\ &&  
\psi_2\left(x_2,t\right)  = \Phi_2\left(x_2-\frac{gt^2}{2\sqrt{2}}\,,t\right) 
\exp{\left[ \;\,\frac {imgt}{\sqrt {2}\hbar }\;\left( x_2-\frac {gt^2}
{6\sqrt {2}}\right) \right]}.
\label{eq83}
\end{eqnarray}
Thus (\ref{eq82}) takes the form
\begin{equation}
\psi \left(x,X,t\right) =e^{\frac {-imgt}{\hbar }\,\frac{x_1+x_2}{\sqrt {2}}}
\Phi_1\left(x_1+\frac{gt^2}{2\sqrt{2}}\,,t\right) 
\Phi_2\left(x_2-\frac{gt^2}{2\sqrt{2}}\,,-t\right).
\label{eq84}
\end{equation}
But $\frac{x_1+x_2}{\sqrt {2}}=X$ and
\begin{eqnarray} &
\Phi_1\left(x_1+\frac{gt^2}{2\sqrt{2}}\,,t\right) 
\Phi_2\left(x_2-\frac{gt^2}{2\sqrt{2}}\,,-t\right)=& \label{eq85} \\ & 
e^{\frac{i}{\hbar}\,\frac{gt^2}{2}\,\frac{\hat p_1-\hat p_2}{\sqrt{2}}}\Phi_1\left(x_1,t
\right) \Phi_2\left(x_2,-t\right)= 
e^{\frac{i}{\hbar}\,\frac{gt^2}{2}\,\hat P}\Phi(x,X,t)=
\Phi\left (x+\frac{gt^2}{2},X,t\right ), & \nonumber
\end{eqnarray}
and we reproduce again (\ref{eq71}). ``I have said it thrice: What I tell you 
three times is true'' \cite{61}.

\section{Propagators in quantum and KvN mechanics}
Green functions or propagators are well-know useful tools in physics. The
propagator $G\left( x_f,t_f; x_i,t_i\right)$ is the probability amplitude for 
finding the particle at the position $x_f$ at time $t_f$, if at time $t_i$ it 
was at the position $x_i$. It can be calculated in a variety of ways, 
especially for Lagrangians which are, at most, quadratic in $x$ 
\cite{P1,P2,P3,P4,P5}. For this restricted class of problems, we can use the 
Fujiwara's formula 
\cite{P1}
\begin{equation}
G\left( x_f,t_f;x_i,t_i\right)=\sqrt{\frac{i}{2\pi\hbar}\,\frac{\partial^2
S\left( x_f,t_f;x_i,t_i\right)}{\partial x_i \partial x_f}}\exp{\left [\frac{i}
{\hbar}S\left( x_f,t_f;\,x_i,t_i\right)\right ]},
\label{eq86}
\end{equation}
where $S\left( x_f,t_f;\,x_i,t_i\right)$ is the classical action. 

For a free particle,
\begin{equation}
S\left(x_f,t_f;\,x_i,t_i\right)=\frac{mv^2}{2}(t_f-t_i)=\frac{m}{2}\,
\frac{(x_f-x_i)^2}{t_f-t_i},
\label{eq87}
\end{equation}
and (\ref{eq86}) gives the corresponding propagator
\begin{equation}
G_0\left( x_f,t_f; x_i,t_i\right)=\sqrt{ \frac {m}{2\pi i \hbar \left( 
t_f-t_i\right) }} \exp {\left [\frac{im\left( x_f-x_i\right)^{2}}{2\hbar 
\left(t_f-t_i\right)}\right ]}.
\label{eq88}
\end{equation}
In the case of a freely falling particle, the solution of the equation of 
motion $\ddot x=-g$ with the boundary conditions $x(t_i)=x_i$ and $x(t_f)=
x_f$ can be written in the following form \cite{62}
\begin{equation}
x-x_0=-\frac{1}{2}\,g(t-t_0)^2,
\label{eq89}
\end{equation}
where
\begin{eqnarray} &&
t_0=\frac{t_i+t_f}{2}+\frac{1}{g}\,\frac{x_f-x_i}{t_f-t_i},\nonumber \\ &&
x_0=\frac{x_i+x_f}{2}+\frac{g}{2}\left [\left(\frac{t_f-t_i}{2}\right)^2+
\frac{1}{g^2}\left(\frac{x_f-x_i}{t_f-t_i}\right)^2\right].
\label{eq90}
\end{eqnarray}
Eq.(\ref{eq89}) reveals a rarely noted  property of uniqueness of free fall:
each solution of $\ddot x=-g$ can be obtained by a spacetime translation
of the primitive solution $x=-\frac{1}{2}gt^2$. Neither a free particle
(relations (\ref{eq90}) are singular at $g=0$), nor a harmonic oscillator 
share this interesting property \cite{62}.

From (\ref{eq89}), $\dot x=-g(t-t_0)$. Thus $\frac{m}{2}\dot{x}^2=
-mg(x-x_0)=\frac{1}{2}mg^2(t-t_0)^2$, and the action integral equals to
\begin{eqnarray} &&
S=\int\limits_{t_i}^{t_f}\left[\frac{m}{2}\dot{x}^2-mgx\right]dt=
\int\limits_{t_i}^{t_f}\left[\frac{m}{2}\dot{x}^2-mg(x-x_0)-mgx_0\right]dt=
\nonumber \\ && \frac{1}{3}mg^2\left [(t_f-t_0)^3-(t_i-t_0)^3\right ]-mgx_0
\left (t_f-t_i\right ).
\label{eq91}
\end{eqnarray}
Substituting $t_0$ and $x_0$ from (\ref{eq90}), we finally obtain
\begin{equation}
S(x_f,t_f; x_i,t_i)=\frac {m\left(x_f-x_i\right)^2 }{2\left(t_f-t_i\right)}-
\frac {1}{2}mg\left( x_i+x_f\right) \left(t_f-t_i\right) -\frac {1}{24}mg^2
\left(t_f-t_i\right)^3,
\label{eq92}
\end{equation}
and then the Fujiwara's formula (\ref{eq86}) gives the corresponding
propagator:
\begin{eqnarray} &
G\left(x_f,t_f; x_i,t_i\right)=\sqrt{\frac {m}{2\pi i\hbar \left(t_f-t_i
\right)}}\times & \nonumber \\ &
\exp{\left \{ \frac {i}{\hbar }\left[ \frac {m\left( x_f-x_i\right)^2 }
{2\left( t_f-t_i\right) }-\frac {1}{2}mg\left( x_i+x_f\right) \left( t_f-t_i
\right)-\frac {1}{24}mg^2\left(t_f-t_i\right)^3\right] \right \}}. &
\label{eq93}
\end{eqnarray}
One can expect from the equivalence principle that $G$ and $G_0$ propagators
are connected to each other in a simple way. This is indeed the case. From 
(\ref{eq32}), $\psi(x,0)=\phi(x,0)$. Therefore,
\begin{equation}
\psi(x,t)=\int\limits_{-\infty}^\infty G(x,t;y,0)\,\psi(y,0)dy=
\int\limits_{-\infty}^\infty G(x,t;y,0)\,\phi(y,0)dy.
\label{eq94}
\end{equation}
On the other hand, again using (\ref{eq32}), we can write
\begin{eqnarray} &
\psi \left( x,t\right) = e^{-\frac {imgt}{\hbar }\left( x+
\frac {gt^{2}}{6}\right)}\,\phi \left( x+\frac {gt^{2}}{2},t\right)=&
\nonumber \\ &
 e^{-\frac {imgt}{\hbar }\left( x+\frac {gt^{2}}{6}\right)}\int\limits_{-\infty}^\infty 
G_0\left (x+\frac{1}{2}gt^2,t;y,0\right )\,\phi(y,0)dy. &
\label{eq95}
\end{eqnarray}
Comparing (\ref{eq95}) and (\ref{eq94}), we see that \cite{P5}
\begin{equation}
G(x_f,t_f;x_i,0)=e^{-\frac {i}{\hbar }mgt_f\left( x_f+\frac {gt_f^{2}}{6}\right)}
G_0\left (x_f+\frac{1}{2}gt_f^2,t_f;x_i,0\right ).
\label{eq96}
\end{equation}
Because of translational invariance in time, (\ref{eq96}) can be written in
a slightly more general form
\begin{equation}
G(x_f,t_f;x_i,t_i)=e^{-\frac {i}{\hbar }mg(t_f-t_i)\left( x_f+\frac {g(t_f-t_i)^{2}}{6}
\right)}G_0\left (x_f+\frac{1}{2}g(t_f-t_i)^2,t_f;x_i,t_i\right ).
\label{eq97}
\end{equation}
The validity of this relationship can be easily verified using (\ref{eq88}) 
and (\ref{eq93}).

To find propagators in KvN mechanics, the simplest way is to use the canonical
transformation (\ref{eq80}) (with unit Jacobian) and the relation (\ref{eq82}).
Then, on the one hand, we will have 
\begin{eqnarray} &&
\psi\left(x_f,X_f,t_f\right )=\iint G^{(KvN)}\left (x_f,X_f,t_f; x_i,X_i,t_i
\right)\psi\left(x_i,X_i,t_i\right )\,dx_idX_i= \nonumber \\ &&
\iint G^{(KvN)}\left (x_f,X_f,t_f; x_i,X_i,t_i
\right)\psi_1\left (x_{1i},t_i\right )\psi_2\left (x_{2i},-t_i\right )\,
dx_{1i}dx_{2i}.
\label{eq98}
\end{eqnarray}
On the other hand, 
\begin{eqnarray} &
\psi\left(x_f,X_f,t_f\right )=\psi_1\left (x_{1f},t_f\right )\psi_2\left (x_{2f},
-t_f\right )=& \label{eq99} \\ & \iint
G(x_{1f},t_f; x_{1i},t_i)G(x_{2f},-t_f; x_{2i},-t_i)
\psi_1\left (x_{1i},t_i\right )\psi_2\left (x_{2i},-t_i\right )\,
dx_{1i}dx_{2i}. &
\nonumber
\end{eqnarray}
Comparing, we see that
\begin{equation}
G^{(KvN)}\left (x_f,X_f,t_f; x_i,X_i,t_i\right)=G(x_{1f},t_f; x_{1i},t_i)
G(x_{2f},-t_f; x_{2i},-t_i).
\label{eq100}
\end{equation}
Since $(x_{1f}-x_{1i})^2+(x_{2f}-x_{2i})^2=(x_{1f}-x_{2f}-x_{1i}+x_{2i})
(x_{1f}+x_{2f}-x_{1i}-x_{2i})=2(x_f-x_i)(X_f-X_i)$, from (\ref{eq100}) we obtain 
a free KvN-propagator
\begin{equation}
G^{(KvN)}_0\left(x_f,X_f,t_f; x_i,X_i,t_i\right) =\frac {m}{2\pi\hbar 
(t_f-t_i)} \exp{\left [\frac {im\left(x_f-x_i\right) \left(X_f-X_i\right)}
{\hbar(t_f-t_i)}\right ]}.
\label{eq101}
\end{equation}
Analogously, from (\ref{eq93}) and (\ref{eq100}), and remembering that the 
gravitational acceleration terms are $g/\sqrt{2}$ and  $-g/\sqrt{2}$ for
$x_1$ and $x_2$ respectively, we get a KvN-propagator in the homogeneous
gravitational field: 
\begin{eqnarray} &
G^{(KvN)}\left(x_f,X_f,t_f; x_i,X_i,t_i\right) = & \label{eq102} \\ &
\frac {m}{2\pi\hbar (t_f-t_i)} \exp{ \left[ \frac {i}{\hbar }\left\{ 
\frac {m}{t_f-t_i}\left(x_f-x_i\right) \left(X_f-X_i\right) -\frac {mg}{2}
\left(X_f+X_i\right) (t_f-t_i)\right\} \right] }. &
\nonumber
\end{eqnarray}
As in the quantum case, there is a relation between the free KvN propagator
and the KvN propagator in a uniform gravitational field, as follows from the 
principle of equivalence. To obtain this relation, we will disentangle 
noncommutative operators in the expression for the KvN propagator in a uniform 
gravitational field \cite{P2}
\begin{equation}
G^{(KvN)}\left(x_f,X_f,t_f; x_i,X_i,t_i\right)=\langle x_f,X_f|e^{-\frac{i}{\hbar}
\left (\frac{\hat p\hat P}{m}+mg\hat X\right)\left (t_f-t_i\right )}|x_i,X_i
\rangle .
\label{eq103}
\end{equation}
Namely, we have (compare with (\ref{eq70}))
\begin{equation}
e^{-\frac{i}{\hbar}
\left (\frac{\hat p\hat P}{m}+mg\hat X\right)\left (t_f-t_i\right )}=
e^{\frac{i}{\hbar}\frac{g\left(t_f-t_i\right)^2}{2}\,\hat P}e^{-\frac{i}{\hbar}\,mg\hat X 
\left(t_f-t_i\right )}e^{-\frac{i}{\hbar}\,\frac{\hat p\hat P}{m}\,\left(t_f-t_i\right)}.
\label{eq104}
\end{equation}
But
\begin{eqnarray} &&
\langle x_f,X_f|e^{\frac{i}{\hbar}\frac{g\left(t_f-t_i\right)^2}{2}\,\hat P}=
\langle x_f+\frac{1}{2}g\left(t_f-t_i\right)^2,X_f|,\nonumber\\ &&
\langle x_f,X_f|e^{-\frac{i}{\hbar}\,mg\hat X \left(t_f-t_i\right )}=
e^{-\frac{i}{\hbar}\,mgX_f \left(t_f-t_i\right )}\langle x_f,X_f|,
\label{eq105}
\end{eqnarray}
and
\begin{equation}
G_0^{(KvN)}\left(x_f,X_f,t_f; x_i,X_i,t_i\right)=\langle x_f,X_f|e^{-\frac{i}{\hbar}\,
\frac{\hat p\hat P}{m}\,\left(t_f-t_i\right)}|x_i,X_i\rangle
\label{eq106}
\end{equation}
is the free KvN-propagator. Therefore,  (\ref{eq104}) implies the following 
relationship:
\begin{eqnarray} &
G^{(KvN)}\left(x_f,X_f,t_f; x_i,X_i,t_i\right)=&\nonumber \\ &
e^{-\frac{i}{\hbar}\,mgX_f \left(t_f-
t_i\right )}\,G_0^{(KvN)}\left(x_f+\frac{1}{2}g\left(t_f-t_i\right)^2,X_f,t_f; 
x_i,X_i,t_i\right), &
\label{eq107}
\end{eqnarray}
similar to (\ref{eq97}). The validity of this relation can be easily verified 
using (\ref{eq101}) and (\ref{eq102}).

It is rather instructive to obtain KvN propagator (\ref{eq102}) by Schwinger's
method \cite{P3}. First of all, since the propagator depends only on the 
difference $\tau=t_f-t_i$, and not separately on $t_f$ and $t_i$, let's
introduce slightly more compact notation:
\begin{equation}
G^{(KvN)}\left(x_f,X_f; x_i,X_i;\tau\right)=\langle x_f,X_f|e^{-\frac{i}{\hbar}\,
\hat H\tau}|x_i,X_i\rangle=\langle x_f,X_f,\tau|x_i,X_i,0\rangle.
\label{eq108}
\end{equation}
From (\ref{eq108}), we get a differential equation 
\begin{eqnarray} &&
i\hbar\frac{\partial }{\partial\tau}G^{(KvN)}\left(x_f,X_f; x_i,X_i;\tau\right)=
\langle x_f,X_f|\hat He^{-\frac{i}{\hbar}\,\hat H\tau}|x_i,X_i\rangle=
\label{eq109} \\ &&
\langle x_f,X_f,\tau|\hat H|x_i,X_i,0\rangle=\left\langle x_f,X_f,\tau\left|
m^{-1}\hat p(0)\hat P(0)+mg\hat X(0)\right |x_i,X_i,0\right\rangle.
\nonumber
\end{eqnarray}
In (\ref{eq108}) and (\ref{eq109}) we have introduced notations
\begin{equation}
|x,X,t\rangle=e^{\frac{i}{\hbar}\hat H t}|x,X\rangle,\;\;\;
\hat x(t)=e^{\frac{i}{\hbar}\hat H t}\hat x e^{-\frac{i}{\hbar}\hat H t},
\hat X(t)=e^{\frac{i}{\hbar}\hat H t}\hat X e^{-\frac{i}{\hbar}\hat H t},
\label{eq110}
\end{equation}
and analogously for $\hat p(t)$ and $\hat P(t$). Note that 
$|x,X,t\rangle$ is a common eigenstate of $\hat x(t)$ and $\hat X(t)$. 
Therefore, one can easily calculate the right-hand side of  (\ref{eq109}), 
if $\hat p(0)\hat P(0)$ can be written in terms of $\hat x(\tau)$, 
$\hat X(\tau)$, $\hat x(0)$ and $\hat X(0)$ in such a way that $\hat x(\tau)$
and $\hat X(\tau)$ appear on the left-hand side, while $\hat x(0)$ and 
$\hat X(0)$ appear on the right-hand side. This can be achieved with the help
of Heisenberg equation of motion $i\hbar\,\frac{d\hat A}{dt}=[\hat A,\,
\hat H]$, which gives
\begin{equation}
\frac{d\hat x}{dt}=\frac{\hat p}{m},\;\;\; \frac{d\hat p}{dt}=-mg,\;\;\;
\frac{d\hat X}{dt}=\frac{\hat P}{m},\;\;\; \frac{d\hat P}{dt}=0.
\label{eq111}
\end{equation}
The solutions of these equations that we need are as follows:
\begin{equation}
\hat x(t)=\hat x(0)+\frac{\hat p(0)}{m}\,t-\frac{1}{2}gt^2,\;\;\;
\hat X(t)=\hat X(0)+\frac{\hat P(0)}{m}\,t.
\label{eq112}
\end{equation}
Therefore,
\begin{equation}
\hat p(0)=\frac{m}{\tau}\left[\hat x(\tau)-\hat x(0)+\frac{1}{2}g\tau^2
\right],\;\;\;\hat P(0)=\frac{m}{\tau}\left[\hat X(\tau)-\hat X(0)\right ],
\label{eq113}
\end{equation}
and we get
\begin{eqnarray}  &
\hat H=\frac{1}{m}\,\hat p(0)\hat P(0)+mg\hat X(0)=
\frac{1}{2}mg\left (\hat X(\tau)+\hat X(0)\right )-
\nonumber & \\ & i\hbar\,\frac{1}{\tau}+\frac{m}{\tau^2}\left[ \left . \left .
\left . \left . \hat X(\tau)
\right (\hat x(\tau)-\hat x(0)\right )-\right(\hat x(\tau)-\hat x(0)\right )
\hat X(0)\right ]. &
\label{eq114}
\end{eqnarray}
To get the correct time-ordering of operators in (\ref{eq114}), we have used
the relation 
$\hat x(0)\,\hat X(\tau)=[\hat x(0),\,\hat X(\tau)]+\hat X(\tau)\,\hat x(0)$, 
and
\begin{equation}
[\hat x(0),\,\hat X(\tau)]=[\hat x(0),\,\hat X(0)+\frac{\hat P(0)}{m}\,\tau]=
i\hbar\,\frac{\tau}{m}.
\label{eq115}
\end{equation}
Substituting (\ref{eq114}) into (\ref{eq109}), we get
\begin{equation}
i\hbar\,\frac{\partial G^{(KvN)}}{\partial \tau}=\left [\frac{m}{\tau^2}\left(
x_f-x_i\right)\left(X_f-X_i\right)-i\hbar\,\frac{1}{\tau}+\frac{1}{2}mg\left(
X_f+X_i\right)\right ] G^{(KvN)}.
\label{eq116}
\end{equation}
This differential equation can be easily integrated with the result
\begin{equation}
G^{(KvN)}\left(x_f,X_f; x_i,X_i;\tau\right)=\frac{C}{\tau}\,e^{\frac{i}{\hbar}
\left[\frac{m}{\tau}\left(x_f-x_i\right)\left(X_f-X_i\right)-\frac{mg}{2}
\left(X_f+X_i\right)\tau\right]},
\label{eq117}
\end{equation}
where $C$ is an integration constant. To determine $C$, we use the initial
condition
\begin{equation}
G^{(KvN)}\left(x_f,X_f; x_i,X_i;0\right)=\langle x_f,X_f,0|x_i,X_i,0\rangle=
\delta\left(x_f-x_i\right)\delta\left(X_f-X_i\right).
\label{eq118}
\end{equation}
Using \cite{63}
\begin{equation}
\lim_{\tau\to 0}\sqrt{\frac{1}{\pi\tau}}\,e^{-\frac{1}{\tau}x^2}=\delta(x),
\label{eq119}
\end{equation}
and $(x_f-x_i)(X_f-X_i)=\frac{1}{2}[(x_{1f}-x_{1i})^2-(x_{2f}-x_{2i})^2]$, which
follows from (\ref{eq80}), we get
\begin{equation}
\lim_{\tau\to 0}\frac{1}{\tau}\,e^{\frac{i}{\hbar}\left[\frac{m}{\tau}
\left(x_f-x_i\right)\left(X_f-X_i\right)\right]}=\frac{2\pi\hbar}{m}\,
\delta(x_{2f}-x_{2i})\delta(x_{1f}-x_{1i}).
\label{eq120}
\end{equation}
But, since the Jacobian of the transformation $(x,X)\to (x_1,x_2)$, implied
by (\ref{eq80}), is unity, $\delta(x_{2f}-x_{2i})\delta(x_{1f}-x_{1i})=
\delta(x_2-x_i)\delta(X_f-X_i)$  \cite{63}, and finally we obtain 
$C=\frac{m}{2\pi\hbar}$. Therefore, (\ref{eq117}) is the same as (\ref{eq102}).

\section{Free fall in quantum and KvN mechanics}
One can expect a thorough consideration of the quantum free fall in standard 
textbooks on quantum mechanics, but, surprisingly, this is not so. Some aspects
of this problem was considered in \cite{64,64A}, and one can find a rather 
exhaustive account in unpublished notes \cite{62}.

Here we are interested in the following question: is it possible to obtain 
information about the mass of a freely falling quantum or KvN probe by 
performing coordinate measurements on the probe during the free fall?

It can be argued that a quantitative measure of the amount of information 
about the mass of the probe we can obtain by performing measurements of its 
coordinate $x$ is provided by Fisher's information \cite{65}
\begin{equation}
I_m=\int dx|\psi(x,t)|^2 \left [\frac{\partial}{\partial m}
\ln{|\psi(x,t)|^2}\right ]^2.
\label{eq121}
\end{equation}
Fisher information in general is a measure of the ability to estimate an 
unknown parameter and is an important concept of mathematical statistics
with many physical applications \cite{65A}.

Suppose we are dropping the following initial Gaussian wave packet
\begin{equation}
\psi \left( x,0\right) =\left( \frac {2}{\pi a^2}\right)^{\frac{1}{4}}\exp \left[ 
-\frac {x^2}{a^2}+\frac {i}{\hbar }\,px\right].
\label{eq122}
\end{equation}
In the momentum space
\begin{equation}
\psi(k)=\frac{1}{\sqrt{2\pi\hbar}}\int\limits_{-\infty}^\infty e^{-\frac{i}{\hbar}\,kx}
\,\psi(x,0)dx=\left(\frac{a^2}{2\pi\hbar^2}\right)^{\frac{1}{4}}
\exp{\left[-\frac{a^2}{4\hbar^2}(k-p)^2\right ]}.
\label{eq123}
\end{equation}
Therefore, (\ref{eq122}) describes a Gaussian wave packet centered around zero
in the configuration space and centered around $p$ in the momentum space. To
find $\psi \left( x,t\right)$, we can use (\ref{eq32}) and time-evolved free
wave-packet \cite{66}
\begin{equation}
\phi(x,t)=\frac {1}{\sqrt {\alpha}}\left(\frac {a^2}{8\pi }\right)^{\frac {1}{4}}
\exp{\left[\frac {i}{\hbar }\,p\left( x-\frac {pt}{2m}\right) \right]}
\exp{\left[-\frac {1}{4\alpha }\left( x-\dfrac {pt}{m}\right)^2\right]},
\label{eq124}
\end{equation}   
where $\alpha=\dfrac {a^{2}}{4}+\dfrac {i\hbar t}{2m} $. Then 
\begin{equation}
|\psi(x,t)|^2=\left|\phi\left (x+\frac{1}{2}gt^2,\,t\right)\right|^2,
\label{eq125}
\end{equation}
with \cite{66}
\begin{equation}
|\phi(x,t)|^2=\frac{1}{\sqrt{2\pi}\Delta(t)}\exp{\left [-\frac{\left(x-
\frac{pt}{m}\right)^2}{2[\Delta(t)]^2}\right ]},\;\;\;\Delta(t)=\frac{a}{2}
\sqrt{1+\frac{4\hbar^2t^2}{m^2a^4}}.
\label{eq126}
\end{equation}
First of all note that, because of (\ref{eq125}), the expectation value of 
$x$ turns out to be
\begin{equation}
\langle x\rangle =\int ^{\infty }_{-\infty }x\left|\psi\left(x,t\right)\right|^2dx
=\int ^{\infty }_{-\infty }\left(y-\frac{1}{2}gt^2\right)\left|\phi\left(y,t\right)
\right|^2dy=\frac {pt}{m}-\frac {1}{2}gt^{2}.
\label{eq127}
\end{equation}
Therefore the wave packet (\ref{eq122}) indeed describes a freely falling 
quantum probe (with initial momentum $p$).

The second consequence of (\ref{eq125}) is that the Fisher information 
related to free fall is the same as for a free particle (we can make a change
of integration variable from $x$ to $y=x+\frac{1}{2}gt^2$ in (\ref{eq121})).
Using
\begin{equation}
\frac{\partial \Delta(t)}{\partial m}=-\frac{\hbar^2t^2}{a^2m^3}\,\frac{1}
{\Delta(t)},
\label{eq128}
\end{equation}
and
\begin{equation}
\frac{\partial \ln{|\phi(x,t)|^2}}{\partial m}=\frac{\hbar^2t^2}{a^2m^3}\left (
\frac{1}{\Delta(t)^2}-\frac{\left (x-\frac{pt}{m}\right)^2}{\Delta(t)^4}
\right )-\frac{x-\frac{pt}{m}}{\Delta(t)^2}\,\frac{pt}{m^2},
\label{eq129}
\end{equation}
it is straightforward to get
\begin{equation}
I_m=\frac{1}{m^2}\left [\frac{2}{\left[1+\left(\frac{ma^2}{2\hbar t}\right)^2
\right]^2}+\left(\frac{pa}{\hbar}\right)^2\frac{1}{1+\left(\frac{ma^2}
{2\hbar t}\right)^2}\right].
\label{eq130}
\end{equation}
As we see, when $t\to\infty$, the Fisher information is saturated to a constant 
value, which means that by monitoring a freely-falling quantum probe, one 
cannot arbitrarily improve the accuracy of determining its mass \cite{65}.
Anyway, this Fisher information doesn't depend on $g$. A homogeneous 
gravitational field does not allow obtaining more information about the mass 
of a freely falling quantum probe than is possible using a free quantum probe.
This constitutes a possible formulation of the weak principle of equivalence
in the quantum realm \cite{65}.

Now consider a KvN particle. In the role of the initial wave packet, we will
take a double Gaussian
\begin{equation}
\psi(x,p,0)=\sqrt{\frac{2}{\pi ab}}\exp{\left[-\frac{x^2}{a^2}-\frac{(p-p_0)^2}
{b^2}\right ]}.
\label{eq131}
\end{equation}
This wave function is in the $(x,p)$-representation where its physical meaning
is especially clear. As we already know, the Koopman Koopman-von Neumann
mechanics is the most quantum-like in the $(x,X)$-representation. In this
representation (\ref{eq131}) takes the form 
\begin{eqnarray} &
\psi(x,X,0)=\frac{1}{\sqrt{2\pi\hbar}}\int\limits_{-\infty}^\infty e^{\frac{i}{\hbar}
pX}\psi(x,p,0)dp= & \nonumber \\ &
\sqrt{\frac {b}{a\pi\hbar }}\exp{\left [-\frac{x^2}{a^2}-
\frac {X^2b^2}{4\hbar^2}+\frac{i}{\hbar}p_0X\right]}. &
\label{eq132}
\end{eqnarray}
In the homogeneous gravitational field, time-evolved wave function 
$\psi(x,X,t)$ can be found by using (\ref{eq71}) and the time-evolved free
wave packet. To find the later, it is convenient first to find a time-evolved 
free wave packet in the $(x,p)$-representation. Free propagator in this 
representation has the following form \cite{67,68} 
\begin{eqnarray} &&
G_0^{(KvN)}(x_f,p_f;x_i,p_i;\tau)=\langle x_f,p_f|e^{-\frac{i}{\hbar}\,\frac{
\hat P \hat p}{m}\,\tau}|x_i,p_i\rangle= \nonumber \\ && 
\left \langle x_f,p_f\left |x_i+\frac{p_i}{m}\tau,\,p_i\right. \right \rangle=
\delta\left(x_f-x_i-\frac{p_i}{m}\tau\right)\delta(p_f-p_i). 
\label{eq133}
\end{eqnarray} 
Therefore, the time-evolved form of (\ref{eq131}) under free evolution is 
\cite{67}
\begin{equation}
\phi(x,p,t)=\sqrt{\frac{2}{\pi ab}}\exp{\left[-\frac{1}{a^2}\left(x-
\frac{p}{m}t\right)^2-\frac{(p-p_0)^2}{b^2}\right ]}.
\label{eq134}
\end{equation}
Then
\begin{eqnarray} &
\phi(x,X,t)=\frac{1}{\sqrt{2\pi\hbar}}\int\limits_{-\infty}^\infty 
e^{\frac{i}{\hbar} pX}\psi(x,p,t)dp= & \nonumber \\ &
\sqrt{\frac{m^2ab}{\pi\hbar(m^2a^2+b^2t^2)}}
\exp{\left[-\frac{x^2}{a^2}-\frac{p^2_0}{b^2}-\frac{\left( 
ib^2tx+ip_0ma^2-\frac{ma^2b^2}{2\hbar}\,X\right)^2}{a^2b^2
\left( m^2a^2+t^2b^2\right)}\right]}, &
\label{eq135}
\end{eqnarray}
and a simple calculation shows that
\begin{equation}
\left |\phi(x,X,t)\right |^2=  \frac{ab}{4\pi\hbar \Delta^2(t)}
\exp{\left[-\frac{1}{2\Delta^2(t)}\left [\left (x-\frac{p_0}{m}\,t\right )^2+
\frac{a^2b^2}{4\hbar^2}\,X^2\right ]\right ]},
\label{eq136}
\end{equation}
where
\begin{equation}
\Delta(t)=\frac{a}{2}\sqrt{1+\frac{b^2t^2}{m^2a^2}}.
\label{eq137}
\end{equation}
Using (\ref{eq71}) and (\ref{eq136}), we can easily find that the expectation 
value of $x$ in the state $\psi(x,X,t)$ is
\begin{equation}
\langle x\rangle =\int \limits^{\infty }_{-\infty } \int\limits ^{\infty }_{-\infty }x
\left| \psi \left( x,X,t\right) \right| ^{2}dx \, dX=\frac {p_{0}t}{m}-
\frac {1}{2}gt^{2}.
\label{eq138}
\end{equation}
Since $X$ is a hidden variable, when calculating Fisher information,  
instead of $|\psi(x,X,t)|^2$ we should use the following probability
density
\begin{equation}
\rho(x,t)=\int\limits_{-\infty}^\infty |\psi(x,X,t)|^2dX=\int\limits_{-\infty}^
\infty \left |\phi\left(x+\frac{gt^2}{2},X,t\right)\right |^2dX.
\label{eq139}
\end{equation}
But
\begin{equation}
I_m=\int\limits_{-\infty}^\infty \rho(x,t)\left [\frac{\partial}{\partial m}
\ln{\rho(x,t)}\right ]^2dx=\int\limits_{-\infty}^\infty \rho(y,t)\left [
\frac{\partial}{\partial m}\ln{\rho(y,t)}\right ]^2dy,
\label{eq140}
\end{equation}
where $y=x+\frac{gt^2}{2}$, and therefore 
\begin{equation}
\rho(y,t)=\int\limits_{-\infty}^\infty |\phi(y,X,t)|^2dX=\frac{1}
{\sqrt{2\pi }\Delta(t)}\exp{\left[-\frac{1}{2\Delta^2(t)}
\left (y-\frac{p_0}{m}\,t\right )^2\right ]}.
\label{eq141}
\end{equation}
Surprisingly enough, (\ref{eq126}) turns out to be a special case
of  (\ref{eq141}) with $b=\frac{2\hbar}{a}$, as (\ref{eq123}) 
indicates. Therefore, the corresponding Fisher information can
be obtained from (\ref{eq130}) by substitution $2\hbar=ab$:
\begin{equation}
I_m=\frac{1}{m^2}\left[\frac{2}{\left[1+\left(\frac{ma}{bt}\right)^2
\right ]^2}+\left(\frac{2p_0}{b}\right)^2\frac{1}{1+\left(\frac{ma}
{bt}\right)^2}\right ].
\label{eq142}
\end{equation}
The behaviour of $I_m$, if $b\ne 0$, is similar to the quantum case.
Note that $I_m=0$ if $p_0=b=0$, but $a\ne 0$. Therefore, if an
assembly of identical classical particles with strictly zero initial
momentum, but some dispersion in initial positions, is dropped in
the uniform gravitational field, we cannot determine the mass of
the particles by observing evolution of their positions in time.
However, if $p_0\ne 0$, $a\ne 0$, but $b=0$, we will have
\begin{equation}
I_m=\frac{4p_0^2t^2}{m^4a^2}.
\label{eq143}
\end{equation}
Therefore, if we know precisely the initial momentum of particles,
observing their positions will allow us to find their mass, and our
ability to do this more precisely will increase quadratically in time.

If in addition $a\to 0$, then $\Delta(t)\to 0$ and it follows from
(\ref{eq119}) and (\ref{eq141})  that $\rho(x,t)=\delta\left (x+
\frac{gt^2}{2}-\frac{p_0}{m}\,t\right )$ that describes a classical
particle (not an assembly of identical particles) dropped in a
uniform gravitational field with some initial momentum $p_0$.
In this case, the Fisher information (\ref{eq143}) diverges: if there
is no uncertainty in the initial position of the particle, a single
measurement of its position at some later time will be sufficient
to find its initial velocity and hence the mass, if the initial
momentum is known.

We can use  (\ref{eq134}) to calculate Fisher information in
the case, when we perform momentum measurements in
addition to the coordinate measurements. The result is
similar to (\ref{eq143}) (and coincides to it when $b=0$):
\begin{equation}
I_m=\frac{t^2}{m^4a^2}\left [b^2+4p_0^2\right ].
\label{eq144}
\end{equation}

\section{Concluding remarks}
Implementation of Einstein's principle of equivalence in KvN
mechanics, as shown in this article, is strikingly similar to its implementation
in quantum mechanics. Therefore, the belief that is sometimes found in the 
literature that the mass dependence of a quantum-mechanical wave function 
implies that the weak principle of equivalence is violated in quantum 
mechanics is erroneous: locally, in the limit of a uniform field, the weak 
principle of equivalence does hold in quantum mechanics in the same way as
in classical mechanics, if the latter is formulated in the language of 
Hilbert space states and operators.

The independence of the Fisher information $I_m$ from the magnitude of the
homogeneous gravitational field is a formulation of Einstein's principle of 
equivalence, which  is universally applicable in both KvN and in quantum 
mechanics.

If the gravitational field is not homogeneous, Fisher information in quantum
mechanics depends on curvature (tidal effects) and quantum wave function
can thus imprint nontrivial information about the mass \cite{65}. The same 
applies to KvN mechanics, although we did not address this issue in this 
article. This is hardly surprising, since it is well known in general 
relativity that a classical extended body can sense tidal forces and thus  
does not follow a geodesic in general \cite{69}.

Both in KvN mechanics and in quantum mechanics, the wave function in the
freely falling reference frame differs from the free wave function only in
phase. In contrast to quantum mechanics, however, in KvN mechanics
this phase depends on the unobservable hidden variable $X$ and thus
can not lead to  the Colella-Overhauser-Werner type interference effects 
observed in quantum mechanics, since we have no control on the variable $X$ 
and cannot organize the splitting and recombination of the particle beam in 
the $X$-direction \footnote{Planck-scale quantum gravity effects can destroy 
unobservability of hidden variables and thus KvN phase factors can potentially 
also lead to some physical effects, but the expected magnitude of such effects 
is vanishingly small \cite{57}.}.

\appendix
\section{Free fall: a historical perspective}
When it comes to remembering Galileo, the first glimpse that most people will 
have is the portrait where he is shown to drop objects of various masses from 
the Leaning tower of Pisa \cite{1}. There is a good reason to believe that this
is merely a legend, and in fact, Galileo never did this experiment:
due to air resistance, such an experiment, if performed, clearly shows that
heavy objects fall faster than light objects \cite{2,3,4}, in full 
confirmation of the views of Aristotle (in fact, it can be proved \cite{4} that
in arbitrary resistive medium the fall time decreases with mass). For example,
in a reconstruction of the Galileo's alleged Leaning Tower of Pisa experiment
a 16-pound shot and a softball were dropped from the tower approximately from 
the 200 feet height, and the shot reached the ground 20-30 feet ahead of the 
softball \cite{2}. 

However, it appears that Galileo did indeed perform some other experiments 
skillfully using inclined planes to dilute the gravitational acceleration,
thereby reducing speeds, as well as the associated air resistance involved in 
the experiments. Although there are grave doubts among professional historians 
of science whether Galileo ever performed any significant experiments at all
\cite{5}, in the case of experiments with an inclined plane, it was at least 
demonstrated that Galileo's alleged experimental results can be reproduced
using devices that are very similar to those described by Galileo \cite{6,7}.
    
One way or another, regardless of the role that real experiments played in 
Galileo's conclusions, these conclusions were quite interesting. Galileo 
asserted that the trajectory of a dropped mass, when it falls under the action 
of gravity, and its time of descent to the ground depend only on its initial 
position and velocity, and is independent of its weight. He argues several
times in his writings that this property of falling bodies is a logical 
necessity that can be inferred from a certain thought experiment. However,
there is a flaw in his argument that becomes especially obvious if we apply
this argument to electric force, where it gives a clearly incorrect conclusion
\cite{8}. In fact, Galileo's universality of free fall is an empirical fact, 
and as such it can be refuted by experiment. For example, Lunar Laser Ranging 
data allows us to verify that Earth and Moon, despite their different masses 
and compositions, fall to the Sun with accelerations that are equal to each 
other with an accuracy of one part in 10 billion \cite{9}.   

Free fall plays a unique role in the history of science, as it has led to three
major breakthroughs in our understanding of gravity. The first was Galileo's 
discovery of universality of free fall. The second breakthrough, the creation
of Newton's theory of universal gravitation, is also associated with the legend.
The legend says that a falling apple inspired Isaac Newton's discovery of the 
law of gravity. The origin of the story can be traced back to Newton himself 
who spoke about this a few months before his death to his younger friend 
William Stukeley \cite{10,11}, and then told the same story to at least three
other people on various occasions \cite{11}. It is not clear why Newton
attributed such significance to this story. Even if the falling apple really
triggered Newton's musings on the nature of gravitation, the growth of theory
from this seed was rather slow and complicated process influenced by Newton's 
interactions with other contemporary scientists \cite{12,13}. 

\section{Free fall and geometrization of gravity}
In modern terms, the main features of the magnificent edifice of Newtonian 
physics, whose foundations were laid by Newton, are the following \cite{14,15,
16,17}. All material processes (events) are imagined to take place in 
spacetime --- a 4-dimensional, real, smooth manifold $M$. Newtonian absolute 
time $t$ is a smooth map with nonvanishing gradient (defined up to a linear 
transformations $t\to at+b$ with $a>0$) from $M$ into the real line. Newtonian 
time defines the absolute (observer independent) simultaneity relation between 
events of $M$ and thus stratifies $M$: the manifold is partitioned into space 
sections each consisting of a totality of simultaneous events. Each space 
section is assumed to have the structure of a Euclidean three dimensional 
space with a positive definite metric. In such Galilean manifold, one can 
distinguish between spacelike vectors (which lay completely in one space 
section) and future or past directed timelike vectors. A world-line of an 
observer is a timelike curve $C$ in $M$ and it can be used to introduce 
Galilean coordinates in $M$. Namely, for any event $P$, let $O$ be the 
simultaneous event on $C$. Then $t(P)=t(O)$ and spatial coordinates of $P$ are 
just Cartesian coordinates of $P$ in the space section containing $P$ and $O$,
with $O$ as the origin of the Cartesian frame. Coordinates defined by 
different Galilean observers are related by the transformation
\begin{equation}
x^{\prime\, i}=R^i_{\,j}(t)x^j+\xi^i(t),\;\;\;t^\prime=t+\tau,
\label{eq1}
\end{equation}
where $\xi^i$ is an arbitrary time-dependent vector, $\tau$ is a real number 
and $R^i_j$ is a real (time-dependent) orthogonal matrix. These transformations
form the kinematical (Leibnizian in another terminology \cite{18}) group 
${\cal K}$. We assume that units for time and length are the same for all 
Galilean observers and therefore the above mentioned linear ambiguity in the 
definition of the Newtonian absolute time is reduced just to the freedom  
(reflected by the presence of $\tau$ in (\ref{eq1})) of Galilean observers to 
choose different origins of their time axis.

Singling out a particular class of world-lines as standard motions, in such a 
way that there is one and only one standard world line through each event and 
time-like direction at that event, then determines the dynamics. Namely, 
standard motions are considered as force-free. For all other motions, forces 
are defined in terms of accelerations relative to the standard motions. The 
generalized Newton's second equation has the form
\begin{equation}
m[\ddot{\vec{x}}-\vec{a}(\vec{x},\dot{\vec{x}})]=\vec{F},
\label{eq2}
\end{equation}
where $\vec{F}$ is the force, $\vec{x}(t)$ describes the motion of a particle 
relative to some Galilean observer, and $\vec{a}(\vec{x},\dot\vec{x})$ is the
the acceleration due to a standard motion of the observer himself.

The Newtonian way of introducing a congruence of preferred world-lines (standard
motions) is to postulate the validity of Newton's law of inertia (the first 
law): There exist the special class of Galilean observers (inertial frames of
reference) with respect to which free particles have zero acceleration:
\begin{equation}
\ddot{\vec{x}}=0.
\label{eq3}
\end{equation}
The transformations (\ref{eq1}) to preserve (\ref{eq3}), one should have
\begin{equation}
\dot{R}^i_{\,j}=0,\;\;\;\ddot{\xi}^i=0.
\label{eq4}
\end{equation}
Such kind of transformations constitute a subgroup of the kinematical group
called the Galilei group ${\cal G}\subset {\cal K}$. Since, according to 
(\ref{eq1}) and (\ref{eq4}), Galilean transformations are linear, it is 
meaningful to call two Galilean 4-vectors parallel if they have the same 
components with respect to inertial coordinate systems. Mathematically
Newton's first law defines on Galilean spacetime an integrable symmetric
connection whose components all vanish in an inertial coordinate system.
Timelike geodesics of this connection represent the free motions.

So far so good, but how to empirically determine inertial reference systems?
With some care, one can eliminate known forces acting on the test body,
such as electric and magnetic forces, drag force, and etc. However, to 
eliminate gravity is not a simple task because even a neutral test body
will accelerate towards massive objects. If one assumes that a test body,
besides the inertial mass $m_i$, has a gravitational mass $m_g$, then Newton's
second law in inertial reference frame takes the form
\begin{equation}
m_i\ddot{\vec{x}}=m_g\,\vec{g}(\vec{x},t)+\vec{F},
\label{eq5}
\end{equation}
where $\vec{g}$ is the gravitational field strength, and $\vec{F}$ is any 
additional non-gravitational force acting on the test body. In arbitrary 
Galilean reference frame non-inertial (fictitious) forces will appear and
(\ref{eq5}) is replaced by
\begin{equation}
\ddot{\vec{x}}=\frac{m_g}{m_i}\,\vec{g}+\frac{\vec{F}}{m_i}-\vec{a}-
2\vec{\omega}\times\dot{\vec{x}}-\dot{\vec{\omega}}\times\vec{x}-
\vec{\omega}\times(\vec{\omega}\times\vec{x}),
\label{eq6}
\end{equation}
where $\vec{a}$ and $\vec{\omega}$ are, respectively, the acceleration
and the angular velocity of the Galilean laboratory frame relative to some 
inertial frame.
 
Universality of free fall means that the ratio $m_g/m_i$ is a universal 
constant that can be taken equal to unity without loss of generality. Then
(\ref{eq6}) indicates that by observing the motions of particles relative 
to some laboratory frame, we can in principle measure $\vec{g}-\vec{a}$, but
not $\vec{g}$ and $\vec{a}$ separately. Gravity and inertial acceleration
are inseparably mingled. Because of this circumstance, the law of inertia
should be replaced by the weak principle of equivalence: There exist free 
fall motions under which gravity is locally canceled by free fall acceleration.

Free fall motions define a special class of Galilean reference frames, the 
Newtonian frames, and if such motions are taken as the standard motions of 
dynamics, the Newton's second law in the Newtonian frames takes the form
\begin{equation}
m(\ddot{\vec{x}}+\nabla \Phi)=\vec{F},
\label{eq7}
\end{equation}
where $\Phi(\vec{x},t)$ is the gravitational potential related to the 
gravitational field strength by relation $\vec{g}=-\nabla \Phi$.

Comparing (\ref{eq7}) and (\ref{eq6}), we see that Newtonian frames are 
non-rotating Galilean frames (Coriolis and centrifugal accelerations are 
absent in the equation of motion of a test body). Therefore, the 
kinematical transformations (\ref{eq1}) that relate different Newtonian frames 
are singled out by the condition that $R^i_{\,j}$ are time-independent: $\dot{R}
^i_{\,j}=0$. Accordingly, the Newtonian group ${\cal N}$ of such transformations
(it is called Maxwellian in \cite{18}) is intermediate in generality between 
the Galilean group ${\cal G}$ and the kinematical group ${\cal K}$: ${\cal G}
\subset {\cal N}\subset {\cal K}$.

Let's demand equation (\ref{eq7}) to be form-invariant (covariant) with 
respect to Newtonian transformations. Due to vector character of (\ref{eq7}),
it is automatically covariant under rotations given by constant orthogonal
matrix $R^i_j$ provided the gravitational potential is a scalar with respect 
to such transformations. The same is true for pure time translations $t^\prime
=t+\tau$. Therefore, without loss of generality, we can consider special
Newtonian transformations
\begin{equation}
\vec{x}^{\,\prime}=\vec{x}+\vec{\xi}(t),\;\;\;t^\prime=t.
\label{eq8}
\end{equation}
Then $\nabla^\prime=\nabla$, $\ddot{\vec{x}}^{\,\prime}=\ddot{\vec{x}}+
\ddot{\vec{\xi}}$, and from (for zero external force)
$\ddot{\vec{x}}^{\,\prime}-\nabla^\prime\Phi^\prime=\ddot{\vec{x}}-\nabla \Phi=0$
we get
\begin{equation}
\nabla(\Phi^\prime-\Phi)=-\ddot{\vec{\xi}},
\label{eq9}
\end{equation}
which shows that under (\ref{eq8}) the gravitational potential is not a scalar
but transforms according to (note that we should have $\Phi^\prime=\Phi$ when
$\vec{\xi}=0$)
\begin{equation}
\Phi^\prime(\vec{x}^{\,\prime},t^\prime)=\Phi(\vec{x},t)-\ddot{\vec{\xi}}\cdot
\vec{x},\;\;\; \nabla^\prime\Phi^\prime=\nabla \Phi-\ddot{\vec{\xi}}.
\label{eq10}
\end{equation}
Hence the splitting of the left-hand-side of (\ref{eq7}) into an inertial term
$m\ddot{\vec{x}}$ and a gravitational term $m\nabla \Phi$ has no objective 
significance as this splitting is frame-dependent: gravitational fields cannot 
be locally separated from translational-acceleration fields.

Integrable flat connection associated with the Galilean transformations does not
remain invariant under Newtonian transformations (\ref{eq8}). Indeed, using
the transformation law of connection coefficients (see, for example, \cite{19}.
Einstein summation convention is assumed)
\begin{equation}   
\Gamma^{\prime\,a}_{bc}(x)=\frac{\partial x^m}{\partial x^{\prime\,b}}\,
\frac{\partial x^n}{\partial x^{\prime\,c}}\,\frac{\partial x^{\prime\,a}}
{\partial x^l}\,\Gamma^l_{mn}(x)+\frac{\partial x^{\prime\,a}}{\partial x^l}\,
\frac{\partial^2 x^l}{\partial x^{\prime b}\partial x^{\prime c}},
\label{eq11}
\end{equation}
one finds that if the connection coefficients are zero in some Newtonian frame,
they don't remain zero after transformation (\ref{eq8}). Namely, (\ref{eq11})
shows that the nonzero connection coefficients in Newtonian coordinates are
$\Gamma^i_{00}$, $i=1,2,3$, and under (\ref{eq8}) they transform according to
\begin{equation}
\Gamma^{\prime\,i}_{00}=\Gamma^i_{00}-\ddot{\xi}^i.
\label{eq12}
\end{equation}
Comparing with (\ref{eq9}), we see that the connection covariant under 
Newtonian transformations (\ref{eq8}) and  (\ref{eq10}) is
\begin{equation}
\Gamma^i_{00}=\frac{\partial \Phi}{\partial x^i}\equiv \Phi_{,i}\;\;{\mathrm if}
\;i=1,2,3, \;\;\;\;\Gamma^a_{bc}=0\;{\mathrm otherwise}.
\label{eq13}
\end{equation}
For such connection, the free fall equation $\ddot{\vec{x}}+\nabla \Phi=0$
takes the form of geodesic equation
\begin{equation}
\frac{d^2 x^a}{d t^2}+\Gamma^a_{bc}\,\frac{dx^b}{dt}\,\frac{dx^c}{dt}=0.
\label{eq14}
\end{equation}
These geodesics are a generalization of the notion of a "straight line" in the 
presence of gravity. As Edmund Whittaker once remarked ``gravitation simply 
represents a continual effort of the universe to straighten itself out''
\cite{20}. More formally, gravity is a non-integrable, symmetric connection, 
whose geodesics are the free fall trajectories \cite{14}. However, to make 
this last statement accurate, one must also geometrize the Poisson equation, 
which in the usual formulation of Newtonian gravity determines the 
gravitational potential. This can be done in the following way \cite{20,21}.

In Newtonian spacetime, we have two degenerate metrics: temporal metric $t_{ab}$ 
and flat spatial metric $h^{ab}$. They can be motivated as follows \cite{20}.
In Minkowski spacetime, and for coordinates $(t,x,y,z)$, we have covariant
metric tensor $g_{ab}=\mathrm{diag}(1,-1/c^2,-1/c^2,-1/c^2)$ and its inverse
contravariant tensor $g^{ab}=\mathrm{diag}(1,-c^2,-c^2,-c^2)$. When the light 
velocity $c$ goes to infinity, well defined limits $t_{ab}=\lim\limits_
{c\to\infty}g_{ab}$ and $h^{ab}=\lim\limits_{c\to\infty}\left (\frac{-g^{ab}}
{c^2}\right)$ will give, correspondingly, Newtonian temporal and spatial 
metrics. These metrics are assumed to obey orthogonality and compatibility 
conditions   
\begin{equation}
t_{ab}h^{ab}=0,\;\;\;\nabla_a\,t_{bc}=0,\;\;\;\nabla_a\,h^{bc}=0.
\label{eq15}
\end{equation}
The first equation can clearly be motivated by $c\to\infty$ limit of the
relation $g_{ab}(g^{ab}/c^2)=4/c^2$. Note that the compatibility conditions
don't uniquely determine the connection, in contrast to general relativity
where $\nabla_\mu\,g_{\nu\lambda}=0$ metric-compatibility condition uniquely
determines the connection. In this sense, the spacetime of geometrized 
Newtonian gravity is not metric (the connection coefficients are not 
expressible in terms of a metric tensor), but only affine \cite{22}.

From connection coefficients (\ref{eq13}), the Riemann tensor can be 
calculated with the result
\begin{equation}
R^i_{0j0}=\partial_j\Gamma^i_{00}=\frac{\partial^2 \Phi}{\partial x^i
\partial x^j},\;\;\;R^a_{bcd}=0\;{\mathrm otherwise}.
\label{eq16}
\end{equation}
Relations (\ref{eq13}) and (\ref{eq16}) can be expressed in covariant form:
\begin{equation}
\Gamma^a_{bc}=t_{bc}h^{ad}\nabla_d\Phi,\;\;\;R^a_{bcd}=t_{bd}h^{ae}\nabla_e\nabla_c
\Phi.
\label{eq17}
\end{equation}
Then $R_{ab}\equiv R^c_{acb}=t_{ab}h^{ce}\nabla_e\nabla_c\Phi=t_{ab}\nabla^2\Phi$
and the Poisson equation $\nabla^2\Phi=4\pi G\rho$ takes the covariant form
\begin{equation}
R_{ab}=4\pi G\rho t_{ab}.
\label{eq18}
\end{equation}
Note that the gravitational potential $\Phi$ completely disappeared in 
(\ref{eq14}) and (\ref{eq18}) and everything is expressed exclusively through 
purely geometric quantities. Newtonian gravity, like the general theory of 
relativity, can also be considered as a spacetime geometry. At that this
geometry becomes dynamical: ``spacetime tells matter how to move; matter tells 
spacetime how to curve'' \cite{23}.

The purpose of this rather long section was to convince the reader 
that free fall plays a central role in the geometrization of gravity not only 
in the general theory of relativity, but already in Newtonian theory. In fact, 
free fall defines a projective structure in spacetime and, as such, is 
necessary for understanding the foundations of any reasonable theory of 
spacetime and gravity \cite{24}. Then an interesting question arises: 
what does the weak equivalence principle mean in quantum theory?

Of course, the understanding that it was impossible to locally distinguish 
between gravity and acceleration was Einstein's insight, who grandly call it 
the ``most fortunate thought in my life'' \cite{25}. 
 
\section{Accelerated reference frame and gravity in classical mechanics}
We begin this section with a reminder of the meaning of passive and active 
transformations (see, for example, \cite{26}). Passive transformation means 
one goes from one reference frame to another which is displaced, rotated, 
etc. relative to the original one. The system itself remains unchanged.
Under active transformation it is the system under study which is
displaced, rotated, etc. The observer's reference frame remains unchanged.
Therefore, an active transformation is simply a change of coordinates in the 
same reference frame.

An example of a passive transformation is the transition from an inertial 
reference frame to an accelerated reference frame. Imagine two reference 
frames $S$ and $S^\prime$ whose origins coincide at $t=0$. Suppose $S$ is an 
inertial reference frame, while $S^\prime$ experiences a free fall in the 
gravitational field $\vec{g}$ present in the frame $S$. It is clear that
coordinates of some event in frames $S$ and $S^\prime$ are related by relation
(\ref{eq8}) with $\ddot{\vec{\xi}}=\vec{g}$. In particular, if $\vec{g}$
points downwards, opposite to the $x$-axis, and the free fall begins at $t=0$ 
with zero relative velocity, we will have 
\begin{equation}
x^\prime=x+\frac{1}{2}gt^2,\;\;y^\prime=y, \;\;z^\prime=z,\;\;t^\prime=t.
\label{eq19}
\end{equation}
This is the passive transformation we will use throughout this paper.

Note that a solution of the Euler-Lagrange equation of motion, which expresses
the final coordinates $x_f$ through the initial coordinates $x_i$, can be 
considered as an active transformation since both coordinates are calculated
relative to the same reference frame.

In reference frame $S$, the gravitational potential is $\Phi=gx$. Then, 
according to (\ref{eq13}), the only non-zero connection coefficient is 
$\Gamma^x_{00}=g$, and (\ref{eq12}) shows that in the frame $S^\prime$ all
connection coefficients vanish. Therefore, gravity is absent in this frame
in conformity with  Einstein's principle of equivalence.

In the frame $S$, the equation of motion follows from the Lagrangian
${\cal L}_g=\frac{m\dot{x}^2}{2}-mgx$, while in the frame  $S^\prime$ the motion
is free and the corresponding Lagrangian is equal to ${\cal L}_f=\frac{m\dot{x}
^{\prime\,2}}{2}$.  These two Lagrangians are equivalent, as they should be. 
Indeed, expressing $x$ through $x^\prime$ from (\ref{eq19}), we get
\begin{equation}
{\cal L}_g=\frac{1}{2}m\dot{x}^{\prime\,2}+mg^2t^2-mg(x^\prime+\dot{x}^{\prime}t)=
{\cal L}_f+\frac{d\Lambda}{dt},\;\;\Lambda=\frac{1}{3}mg^2t^3-mg x^\prime t.
\label{eq20}
\end{equation}
However, any two Lagrangians that differ by a total time derivative are 
equivalent in the sense that they give the same equations of motion \cite{26}.
This equivalence is a mathematical expression of  Einstein's principle of 
equivalence in classical mechanics.

As it is well known, in the quasiclassical approximation the phase of the 
quantum wave function coincides with the classical action integral along
the classical trajectory \cite{27,28}. Therefore, equation (\ref{eq20}) suggest
that wave functions of a quantum particle in the uniform gravitational field
and a free quantum particle in the freely falling non-inertial reference frame
differ only in phase and this phase difference equals to $\Lambda/\hbar$.

The very same conclusion can be reached from the Hamiltonian perspective. If
${\cal L}_g$ is expressed through $x^\prime$, as in (\ref{eq20}), then the 
corresponding canonical momentum is $p^\prime=\frac{\partial {\cal L}_g}
{\partial \dot{x}^\prime}=m \dot{x}^\prime-mgt$. But (\ref{eq19}) indicates that
$\dot{x}^\prime-gt=\dot{x}$, and therefore $p^\prime=p$. As we see, transition
to the freely falling frame is equivalent to the following canonical 
transformation
\begin{equation}
x^\prime=x+\frac{1}{2}gt^2,\;\;\;p^\prime=p.
\label{eq21}
\end{equation}
This transformation is time dependent and thus the new Hamiltonian is \cite{26}
\begin{equation}
{\cal H}^\prime(x^\prime,p^\prime,t)={\cal H}(x(x^\prime,p^\prime),p(x^\prime,p^\prime)
,t)+\frac{\partial F(x(x^\prime,p^\prime),p^\prime,t)}
{\partial t},
\label{eq22}
\end{equation}
where $F(x,p^\prime,t)$ is the second type Legendre generator of the canonical 
transformation (\ref{eq21}) \cite{26}:
\begin{equation}
p=\frac{\partial F(x,p^\prime,t)}{\partial x},\;\;\;
x^\prime=\frac{\partial F(x,p^\prime,t)}{\partial p^\prime}.
\label{eq23}
\end{equation}  
If we put $x^\prime=x+gt^2/2$ in the second equation of  (\ref{eq23}) and 
integrate, we get $F(x,p^\prime,t)=p^\prime(x+\frac{1}{2}gt^2)+G(x,t)$, and then
the first equation will give $p^\prime=p=\frac{\partial F(x,p^\prime,t)}
{\partial x}=p^\prime+\frac{\partial G(x,t)}{\partial x}$. Therefore, $G(x,t)=
f(t)$, where $f(t)$ is an arbitrary function of time, and the  Legendre 
generator finally takes the form
\begin{equation}
F(x,p^\prime,t)=p^\prime(x+\frac{1}{2}gt^2)+f(t).
\label{eq24}
\end{equation} 
In light of $p^\prime=p=m\dot{x}=m(\dot{x}^\prime-gt)$, we get from (\ref{eq22})
\begin{eqnarray}
{\cal H}^\prime=&\frac{p^{\prime\,2}}{2m}&+mg\left(x^\prime-\frac{1}{2}gt^2\right)+
mgt(\dot{x}^\prime-gt)+\dot{f}(t)= \nonumber \\ &\frac{p^{\prime\,2}}{2m}&+
\frac{d}{dt}(mgx^\prime t)-\frac{3}{2}mg^2t^2+\dot{f}(t).
\label{eq25}
\end{eqnarray}
If now we choose $f(t)=\frac{1}{6}mg^2t^3$, we get
\begin{equation}
{\cal H}^\prime=\frac{p^{\prime\,2}}{2m}-\frac{d\Lambda}{dt}.
\label{eq26}
\end{equation}
This relation shows that in the freely falling reference frame $S^\prime$, 
the Hamiltonian ${\cal H}^\prime$ of a particle in a homogeneous gravitational
field differs from the free Hamiltonian $\frac{p^{\prime\,2}}{2m}$ by a gauge
term $-\frac{d\Lambda}{dt}$. Therefore, again it is expected that the 
corresponding wave functions in frames $S$ and $S^\prime$ differ only by a phase
$\Lambda/\hbar$. As we have seen in the main text, this is indeed true in 
quantum mechanics. 

\section{KvN mechanics: an introductory review}
The one-dimensional Schr\"{o}dinger equation, which describes the dynamics of
a quantum particle in terms of the wave function $\psi\left( x,t\right)$, 
is given by
\begin{equation}
i\hbar \,\frac {\partial \psi\left( x,t\right) }{\partial t} =
-\frac {\hbar ^{2}}{2m}\dfrac {\partial ^{2}\psi\left( x,t\right) }
{\partial x^{2}} +V\left( x \right) \psi \left( x,t\right).
\label{eq38}
\end{equation}
The probability density to find a quantum particle at a particular position 
$x$ at a time $t$ is given by modulus squared of the wave function  
$\rho(x,t)=|\psi(x,t)|^2$. Interestingly, there is a close relationship of 
(\ref{eq38}) with the diffusion equation \cite{43}. For a free particle 
($V=0$), if we define the imaginary time $\tau=it$, then the Schr\"{o}dinger 
equation will look like a diffusion equation (also called the heat equation) 
with a diffusion coefficient $D=\frac{\hbar}{2m}$\footnote{In a remarkable 
paper \cite{44} Nelson proposed a derivation of the Schr\"{o}dinger equation, 
suggesting that a quantum particle is subject to Brownian motion with the 
diffusion coefficient $D=\frac{\hbar}{2m}$.} :
\begin{equation}
\frac {\partial \psi }{\partial \tau}=D\dfrac {\partial ^{2}\psi }
{\partial x^{2}}.
\label{eq39}
\end{equation} 
Let's make the Hopf-Cole transform \cite{45}
\begin{equation}
\psi(x,t) =e^{-\frac{1}{2D}\int\limits_0^x u(y,t)dy}
\label{eq40}
\end{equation}
in (\ref{eq39}). Then we get 
$$\frac{\partial}{\partial t}\left (\int\limits_0^x u(y,t)dy\right )+
\frac{1}{2}u^2=D\,\frac{\partial u}{\partial x},$$
and taking the partial derivative with respect to $x$, we get the Burgers 
equation for $u(x,t)$:
\begin{equation}
\frac{\partial u}{\partial t}+u\,\frac{\partial u}{\partial x}=
D\,\frac{\partial^2 u}{\partial x^2}.
\label{eq41}
\end{equation}
The Burgers equation is an approximation of the Navier-Stokes equations when
pressure is neglected, but the influence of the nonlinear and viscous terms 
is preserved \cite{46}. In light of similarities between (\ref{eq38}) and
(\ref{eq39}), it is expected that the Schr\"{o}dinger equation can also
have a hydrodynamic formulation. In a seminal paper \cite{47} Madelung
showed that this is indeed true.

Madelung transform is a complex analog of the Hopf-Cole transform 
(\ref{eq40}):
\begin{equation}
\psi(x,t)=e^{\frac{1}{2}\ln{\rho(x,t)}+\frac{i}{\hbar}S(x,t)}=\sqrt{\rho(x,t)}
e^{\frac{i}{\hbar}S(x,t)}.
\label{eq42}
\end{equation}
If we substitute (\ref{eq42}) into the Schr\"{o}dinger equation (\ref{eq38}),
we obtain the following equations after separation of the real and imaginary 
parts:
\begin{eqnarray} &&
\frac {\partial \rho }{\partial t}+\frac {\partial }{\partial x}
\left( \rho v\right) =0, \nonumber \\ &&
\frac {\partial S}{\partial t}+\frac {1}{2m}\left( \frac {\partial S}
{\partial x}\right) ^{2}+V\left( x,t\right) +Q\left( x,t\right) =0,
\label{eq43}
\end{eqnarray}
where 
\begin{equation}
Q(x,t)=-\frac {\hbar^2 }{2m\sqrt {\rho}}\left( \frac {\partial ^{2}\sqrt 
{\rho }}{\partial x^{2}}\right),\;\;\; v=\frac{1}{m}\frac {\partial S}
{\partial x}.
\label{eq44}
\end{equation}
In the three-dimensional case, the corresponding equations are \cite{47}
\begin{eqnarray} &&
\frac{\partial \rho }{\partial t}+\nabla\cdot\left( \rho \vec{v}\right) =0, 
\nonumber \\ &&
\frac {\partial S}{\partial t}+\frac {1}{2m}\left(\nabla S\right)^2+
V\left( x,t\right) +Q\left( x,t\right) =0,
\label{eq45}
\end{eqnarray}
and
\begin{equation}
Q=-\frac {\hbar^2 }{2m\sqrt {\rho}}\,\nabla^2 \sqrt {\rho}=
-\frac {\hbar^2 }{2m}\left[\left(\nabla\ln{\sqrt{\rho}}\right)^2+\nabla^2
\ln{\sqrt{\rho}}\right],
\;\;\; v=\frac{1}{m}\nabla S.
\label{eq46}
\end{equation}
Using $\epsilon_{ijk}\epsilon_{kmn}=\delta_{im}\delta_{jn}-\delta_{in}\delta_{jm}$,
it is easy to prove that $\vec{v}\times(\nabla\times\vec{v})=\frac{1}{2}\nabla
v^2-(\vec{v}\cdot\nabla)\vec{v}$. But $\nabla\times\vec{v}=0$ since $\vec{v}$
is a gradient. Therefore, if we apply $\frac{1}{m}\nabla$ operator to both 
sides of the second equation of (\ref{eq45}), we get 
\begin{equation}
\frac{\partial \vec{v}}{\partial t}+(\vec{v}\cdot\nabla)\vec{v}=-\frac{1}{m}
\nabla (V+Q).
\label{eq47}
\end{equation} 
This last equation has close analogy with an irrotational flow of a  perfect 
fluid of density $\rho$ in a gravitational field $\vec{g}$ described by Euler 
equation 
\cite{48}
\begin{equation}
\frac{\partial \vec{v}}{\partial t}+(\vec{v}\cdot\nabla)\vec{v}=\vec{g}-
\frac{\nabla p}{\rho},
\label{eq48}
\end{equation} 
where $p$ is the pressure.

Indeed, using $\nabla_j\ln{\sqrt{\rho}}=\nabla_j\rho/(2\rho)$, we get
\begin{eqnarray} &&
\left(-\frac{2m}{\hbar^2}\right)\nabla_iQ=
\left[2(\nabla_j\ln{\sqrt{\rho}})\nabla_i\nabla_j
\ln{\sqrt{\rho}}+\nabla_i\nabla_j\nabla_j\ln{\sqrt{\rho}}\right]= 
\nonumber \\ && 
\left .  \left . \frac{1}{\rho}\right [(\nabla_j\rho)\nabla_i\nabla_j
\ln{\sqrt{\rho}}+\rho\nabla_j\nabla_i\nabla_j\ln{\sqrt{\rho}}\right]=
\frac{1}{\rho}\,\nabla_j\left(\rho\nabla_i\nabla_j\ln{\sqrt{\rho}}\right). 
\label{eq49}
\end{eqnarray}
Therefore,
\begin{equation}
\frac{1}{m}\nabla_iQ=\frac{1}{m\rho}\nabla_j\sigma_{ij},\;\;\;
\sigma_{ij}=-\frac{\hbar^2}{2m}\rho\,\nabla_i\nabla_j\ln{\sqrt{\rho}}.
\label{eq50}
\end{equation}
As a result, comparing (\ref{eq47}) and (\ref{eq48}), we see that the Madelung
equations (\ref{eq45}) can be considered as a description of the hydrodynamics 
of a specific fluid with mass density $m\rho$, in which the pressure tensor 
$p\delta_{ij}$ is replaced by the tensor $\sigma_{ij}$ \cite{49,50}.

Madelung representation of the Schr\"{o}dinger equation proved to be useful 
and is extensively used in the description of physical phenomena like 
superfluidity, Bose-Einstein condensation, quantum plasmas, to name a few. 
Oddly enough, Madelung's contemporaries paid very little attention to this 
approach, and Pauli even expressed the opinion that it was not very 
interesting \cite{51}. 

It is noteworthy that a fundamental equation of statistical mechanics, the 
Liouville equation
\begin{equation}
\frac {\partial f(\vec{x},\vec{p},t)}{\partial t}=\left\{ H(\vec{x},\vec{p},
t), f(\vec{x},\vec{p},t)\right\},\;\;\;
\left\{ H, f \right\}=\frac{\partial H}{\partial x_i}\,\frac{\partial f}
{\partial p_i}-\frac{\partial H}{\partial p_i}\,\frac{\partial f}
{\partial x_i},
\label{eq51}
\end{equation}
also admits a hydrodynamic interpretation. This can be shown as follows 
\cite{52}.

Let's  make hydrodynamic substitution $ f(\vec{x},\vec{p},t)=\rho(\vec{x},
t)\,\delta(\vec{p}-m\vec{v}(\vec{x},t))$ in the Liouville equation and 
integrate over $\vec{p}$ under the assumption of the usual Hamiltonian 
$H=\vec{p}^{\,2}/(2m)+V(\vec{x})$. Since 
$$\int\frac{\partial V}{\partial x_i}\frac{\partial f}
{\partial p_i}\,d\vec{p}=\rho\frac{\partial V}{\partial x_i}
\int\frac{\partial}{\partial p_i}\delta(\vec{p}-m\vec{v})\,d\vec{p}=0,$$
and
$$\int\frac{p_i}{m}\,\frac{\partial f}{\partial x_i}\,d\vec{p}=\frac{\partial}
{\partial x_i}\left (\rho \int\frac{p_i}{m}\,\delta(\vec{p}-m\vec{v})\,d\vec{p}
\right )=\frac{\partial}{\partial x_i}(\rho v_i),$$
we immediately  obtain  the  continuity equation (the first equation of the 
Euler system (\ref{eq45})).

To obtain Euler's momentum equation, we multiply the Liouville equation by
$p_i$ and integrate over $\vec{p}$. Since 
$$\int\frac{\partial V}{\partial x_j}\frac{\partial f}
{\partial p_j}\,p_i\,d\vec{p}=\rho\,\frac{\partial V}{\partial x_j}
\int p_i\frac{\partial}{\partial p_j}\delta(\vec{p}-m\vec{v})\,d\vec{p}=
-\rho\,\frac{\partial V}{\partial x_i}$$
and
$$\int\frac{p_j}{m}\,\frac{\partial f}{\partial x_j}\,p_i\,d\vec{p}=
\frac{\partial}{\partial x_j}\left (\rho \int\frac{p_j}{m}\,p_i\,
\delta(\vec{p}-m\vec{v})\,d\vec{p}\right )=m\frac{\partial}{\partial x_j}
(\rho\, v_iv_j),$$
we obtain
\begin{equation}
\frac{\partial}{\partial t}(\rho v_i)+\frac{\partial}{\partial x_j}
(\rho v_j v_i)=-\frac{\rho}{m}\,\frac{\partial V}{\partial x_i}.
\label{eq52}
\end{equation}
Because of the continuity equation, (\ref{eq52}) is equivalent to 
(\ref{eq47}) without the quantum potential $Q$. Indeed, the left hand side 
of (\ref{eq52}) equals to 
$$\rho\frac{\partial v_i}{\partial t}+v_i\left(\frac{\partial 
\rho}{\partial t}+\frac{\partial}{\partial x_j}(\rho v_j)\right )+\rho v_j
\frac{\partial v_i}{\partial x_j}=\rho\frac{\partial v_i}{\partial t}+
\rho v_j\frac{\partial v_i}{\partial x_j}.$$
 
Since both the Schr\"{o}dinger equation and the Liouville equation admit 
a hydrodynamic interpretation, the natural question is whether it is possible 
to formulate classical mechanics in terms of a wave function 
$\psi\left(x,p,t\right)$ in phase space. This wave function must be of 
Schr\"{o}dinger type, that is, the following postulates usually attributed to 
quantum theory should be satisfied:
\begin{itemize}
\item The states of a classical mechanical system are represented by 
normalized vectors $|\psi\rangle$ of a complex Hilbert space;
\item Observables are given by self-adjoint operators acting on this Hilbert
space;
\item The expectation value of an observable $\hat A$ in a given state 
$|\psi\rangle$ is $\bar A=\langle\psi|\hat A|\psi\rangle$, where the inner 
product equals to 
\begin{equation}
\langle\psi|\phi\rangle=\int dqdp\, \psi^*(p,q,t)\phi(p,q,t),
\label{eq53}
\end{equation}
\item If $|A\rangle$ is an eigenvector of the observable $\hat A$ with 
eigenvalue $A$, the probability that a measurement of the observer $\hat A$ 
in a given state 
$|\psi\rangle$ yields $A$ is $|\langle A|\psi\rangle|^2$. 
\end{itemize}
Such kind of scenario can be realized as follows \cite{53,54}. The unitary 
operators $\hat U(t)$ that connect state vectors at time $0$ to state vectors 
at time $t$,
\begin{equation}
|\psi(t)\rangle=\hat U(t)|\psi(0)\rangle,
\label{eq54}
\end{equation}
constitute a unitary representation of the one-parameter group of time 
translations.  According to Stone's theorem, $\hat U(t)=e^{-\frac{i}{\hbar}
\hat H}$, where $\hat H$ is Hermitian (a Physicist's proof of Stone's 
theorem can be found in \cite{55}, and some refinements in \cite{56}). 
Then (\ref{eq54}) imply 
\begin{equation}
i\hbar\frac{d}{dt}|\psi(t)\rangle=\hat H|\psi(t)\rangle,
\label{eq55}
\end{equation}
As a fundamental principle determining $\hat H$, we assume that the evolution 
of expectation values of the coordinate and momentum operators is governed by 
the Ehrenfest theorem (a kind of correspondence principle that requires 
Newton's equations for these expectation values)
\begin{equation}
\frac{d}{dt}\langle\hat x\rangle=\langle\frac{\hat p}{m}\rangle,\;\;\;
\frac{d}{dt}\langle\hat p\rangle=\langle-\hat{V}^\prime(\hat x)\rangle,
\label{eq56}
\end{equation} 
while the operators $\hat x$ and $\hat p$ themselves commute:
\begin{equation}  
\left[ \hat {x},\hat {p}\right] =0.
\label{eq57}
\end{equation} 
Note that in order to obtain the operator $\hat{V}^\prime(\hat x)$ we need
to take the derivative of the potential $V(x)$ by treating $x$ as a 
$c$-number and  then converting $x\to \hat x$ in the result \cite{54}.

Because of (\ref{eq55}) and (\ref{eq56}), 
$$i\hbar\frac{d}{dt}\langle\hat x\rangle=i\hbar\langle\psi(t)|\hat x|\psi(t)
\rangle=\langle\psi(t)|[\hat{x},\hat{H}]|\psi(t)\rangle=i\hbar\langle\psi(t)|
\frac{\hat p}{m}|\psi(t)\rangle,$$
and similarly for $i\hbar\frac{d}{dt}\langle\hat p\rangle$. As relations
(\ref{eq56}) must be satisfied  for any state vector $|\psi(t)\rangle$, we get
\begin{equation}  
\frac{i}{\hbar}\left[ \hat {H},\hat {x}\right] =\frac{\hat p}{m},\;\;\;
\frac{i}{\hbar}\left[ \hat {H},\hat {p}\right] =-\hat{V}^\prime(\hat x).
\label{eq58}
\end{equation}
However, in light of (\ref{eq57}), if $\hat H$ is a function of only $\hat x$
and $\hat p$, the commutation relations  (\ref{eq58}) are clearly impossible.
Therefore we introduce two new operators $\hat X$ and  $\hat P$ with 
commutation relations
\begin{equation}
\left[ \hat {X},\hat {p}\right] =i\hbar,\left[ \hat {X},\hat {P}\right] =0,
\left[ \hat {p},\hat {P}\right] =0,\left[ \hat {x},\hat {X}\right] =0,
\left[ \hat {x},\hat {p}\right] =0,\left[ \hat {x},\hat {P}\right]=i\hbar,
\label{eq59}
\end{equation}
and suppose that $\hat H(\hat x,\hat X,\hat p,\hat P)$ depends on these new
operators too. Then
\begin{equation}
\frac{i}{\hbar}\left [\hat H,\hat x\right ]=\frac{\partial \hat H}{
\partial\hat P},\;\;\;\frac{i}{\hbar}\left [\hat H,\hat p\right ]=
-\frac{\partial \hat H}{\partial\hat X},
\label{eq60}
\end{equation}
and comparing with  (\ref{eq58}) we see that
\begin{equation}
\hat H=\frac{\hat{p}\hat{P}}{m}+\hat{V}^\prime(\hat x)\hat{X}+
f(\hat x,\hat p,t),
\label{eq61}
\end{equation}
where $f(\hat x,\hat p,t)$ is an arbitrary real function.

Since $\hat x$ and $\hat p$ are commuting operators, we can use a basis of 
their common eigenstates $|x,p\rangle$ to define the wave function 
$\psi(x,p,t)=\langle x,p|\psi(t)\rangle$. In this $(x,p)$-representation
\begin{equation}
\hat P=-i\hbar\frac{\partial}{\partial x},\;\;\;
\hat X=i\hbar\frac{\partial}{\partial p},\
\label{eq62}
\end{equation}  
and the Schr\"{o}dinger equation (\ref{eq55}) takes the form
\begin{equation}
\frac{\partial \psi(x,p,t)}{\partial t}=\left (V^\prime(x)\frac{\partial}
{\partial p}-\frac{p}{m}\,\frac{\partial}{\partial x}-\frac{i}{\hbar}
f(x,p,t)\right )\psi(x,p,t).
\label{eq63}
\end{equation}
Then the probability density $\rho(x,p,t)=\psi^*(x,p,t)\psi(x,p,t)$ satisfies
the equation
\begin{equation}
\frac{\partial \rho(x,p,t)}{\partial t}=V^\prime(x)\frac{\partial \rho(x,p,t)}
{\partial p}-\frac{p}{m}\,\frac{\partial \rho(x,p,t)}{\partial x}=
\{H_{cl},\rho\},
\label{eq64}
\end{equation}
which is nothing but the Liouville equation (\ref{eq51}) with the classical
Hamiltonian $H_{cl}=\frac{p^2}{2m}+V(x)$.

As we see, the presence of the arbitrary function $f(x,p,t)$ in (\ref{eq61})
does not affect the observed probability density $\rho$ and, thus, it can be
safely ignored. In fact, this gauge freedom in the choice of the Hamiltonian
(\ref{eq61}) (the Liouville operator) is related to the invariance of the KvN
probability density function under the phase transformation of the 
corresponding wave function \cite{57} and reflects the fact that the phase of
the wave function is irrelevant in classical theory.

If $f(x,p,t)=0$, then both $\rho(x,p,t)$ and $\psi(x,p,t)$ satisfy the same 
Liouville equation (\ref{eq64}). This difference between KvN mechanics and 
quantum mechanics is due to the fact that in the $(x,p)$-representation
the  Liouville equation is linear in derivatives.

Interesting perspective on KvN mechanics was given by  Sudarshan \cite{58,59}.
We can consider $(x,X,p,P)$ as the phase space of a quantum system with twice 
as many degrees of freedom than the classical system  $(x,p)$. In the subspace
$(x,p)$, the classical dynamics arises due to the very unusual form of the
quantum Hamiltonian (\ref{eq61}), which is linear in the classically hidden 
variables $X$ and $P$. 

The dynamics of a KvN particle will look more quantum-like in the 
$(x,X)$-representation. In this representation $\hat x$ and $\hat X$ operators 
are diagonal, while
\begin{equation}
\hat P=-i\hbar\frac{\partial}{\partial x},\;\;\;
\hat p=-i\hbar\frac{\partial}{\partial X},\
\label{eq65}
\end{equation}
and the Schr\"{o}dinger equation (\ref{eq55}) takes the form
\begin{equation}
i\hbar\frac{\partial \psi(x,X,t)}{\partial t}=-\frac{\hbar^2}{m}\,
\frac{\partial^2 \psi(x,X,t)}{\partial x \partial X}+X\frac{dV}{dx}\,
\psi(x,X,t).
\label{eq66}
\end{equation}
Hilbert space approach to classical mechanics was pioneered  by Koopman 
\cite{I11} and von Neumann \cite{I12}. A detailed up-to-date pedagogical 
review is given in \cite{60}, where the interested reader can also find  
relevant references regarding contemporary research related to KvN mechanics
(in this regard, see also \cite{57}).

\section*{Acknowledgments}
We are grateful to the anonymous referee for comments that helped to improve 
the presentation of the manuscript. The work of Z.K.S. is supported by the 
Ministry of Education and Science of the Russian Federation.

\end{document}